\documentclass[a4paper,11pt]{article}
\pdfoutput=1 

\usepackage{jheppub}

\usepackage[T1]{fontenc}

\usepackage{amsmath,amssymb,amsfonts, physics}
\usepackage{mathrsfs}
\usepackage{bbm}
\usepackage{slashed}
\usepackage{booktabs}
\usepackage{graphicx}
\usepackage{subcaption}
\usepackage{verbatim}
\usepackage[T1]{fontenc}
\usepackage{placeins}
\usepackage{subcaption}
\usepackage[utf8]{inputenc}
\usepackage[dvipsnames]{xcolor}
\usepackage[normalem]{ulem}
\usepackage{svg}

\usepackage[backend=bibtex,style=numeric-comp, sorting=none]{biblatex}
\bibliography{references}

\definecolor{pink}{RGB}{255,192,203}

\definecolor{green}{RGB}{10, 143, 121}

\title{\boldmath An RG perspective on the (un)naturalness of Starobinsky inflation}

\author[a,b]{Francesca Campani}
\author[a,b]{Francesco Ferrarin}
\author[c]{Serena Giardino}
\author[a,b,d]{Alessia Platania}
\affiliation[a]{Center of Gravity, Niels Bohr Institute, Blegdamsvej 17, DK-2100 Copenhagen \O, DENMARK}
\affiliation[b]{Niels Bohr International Academy, The Niels Bohr Institute, Blegdamsvej 17, DK-2100 Copenhagen \O, DENMARK}
\affiliation[c]{Abdus Salam Centre for Theoretical Physics, Imperial College, London, SW7 2AZ, UK}
\affiliation[d]{Institute of Physics, University of Graz, Universit\"atsplatz 5, 8010 Graz, Austria}

\emailAdd{mzg977@alumni.ku.dk}
\emailAdd{francesco.ferrarin@nbi.ku.dk}
\emailAdd{s.giardino@imperial.ac.uk}
\emailAdd{alessia.platania@nbi.ku.dk}

\abstract{Starobinsky inflation is often regarded as a successful minimal model of the early universe, yet it relies on a large $R^2$-coupling, in apparent tension with the expectation that Wilson coefficients in a gravitational effective field theory are of the same order. We test this expectation with a first-principles renormalization group calculation. We use asymptotically safe gravity as a testing ground to define ultraviolet-complete renormalization group trajectories and to uniquely predict higher-order couplings in terms of the free parameters of the theory---the cosmological constant and the $R^2$-coupling in Planck units. Including higher-curvature operators up to the fourth order, we find that indeed the flow generates dimensionless couplings of comparable size in the ultraviolet, with the irrelevant couplings tied to the relevant ones by linear relations. However, at sub-Planckian energies a canonical-like scaling takes over, and the irrelevant couplings get suppressed by powers of the ratio between the inflationary and Planck scales. This hierarchy yields the suppression of higher-order operators needed for consistency with cosmological data, so that the $R^2$-dominated dynamics emerges as a natural low-energy approximation. At the same time, the higher-order terms required by fundamental physics induce small but calculable deviations from the Starobinsky predictions, which may be worth considering in light of the latest and future DESI and ACT results.} 

\begin{document} 
\maketitle
\flushbottom

\section{Introduction}\label{sect:introduction}

Starobinsky inflation~\cite{Starobinsky:1980te,Mukhanov:1981xt} is arguably the simplest inflationary model, and it has long been in remarkable agreement with cosmological observations. Its action consists of the Einstein-Hilbert term supplemented by a single higher-curvature operator, $R^2$, and is equivalent to a scalar field with a plateau-like potential. The resulting predictions for the scalar spectral index and the tensor-to-scalar ratio fit the Cosmic Microwave Background (CMB) data very well~\cite{Planck:2018jri}. At the same time, it should be stressed that the pure Starobinsky model---in which all higher-curvature couplings are not just suppressed but enforced to be exactly zero---is phenomenological: it is a minimal choice that seems to fit the data, but it lacks a fundamental physics explanation. In particular, higher-derivative operators are expected on general grounds, from any consistent ultraviolet (UV) complete framework for quantum gravity (QG) and matter. On the observational side, the most recent CMB data releases~\cite{AtacamaCosmologyTelescope:2025blo,AtacamaCosmologyTelescope:2025nti}, combined with Baryon Acoustic Oscillation (BAO) data from the Dark Energy Spectroscopic Instrument (DESI)~\cite{DESI:2024mwx,DESI:2025zgx}, mildly disfavor its predictions; see~\cite{McDonough:2025lzo} for an analysis of this tension. This motivates a closer look at corrections to the minimal model, and in particular at the same higher-derivative operators that render the pure Starobinsky Lagrangian fine-tuned. 

In this context, a feature of the Starobinsky action that is rarely scrutinized is its embedding in an effective field theory (EFT) framework. Matching the observed amplitude of scalar perturbations requires the coefficient of the $R^2$ term to be very large, of order $\sim 10^{12}$. From the viewpoint of gravity as an EFT~\cite{Donoghue:1994dn,Burgess:2003jk}, this is puzzling for two reasons. First, there is no symmetry principle that singles out quadratic order: the renormalization group (RG) generates all operators compatible with diffeomorphism invariance, so that a consistent EFT contains higher powers of the Ricci scalar---$R^3$, $R^4$, and so on---as well as operators built out of curvature tensors and their derivatives. As a complementary viewpoint, in the context of the swampland program~\cite{Brennan:2017rbf,Palti:2019pca,Agmon:2022thq,VanRiet:2023pnx,McNamara:2022xkg}, the pure Starobinsky model has been placed in the ``swampland''~\cite{Lust:2023zql,Ketov:2024klm}---the set of inconsistent EFTs. Second, the common lore is that all Wilson coefficients should be of the same order~\cite{Manohar:1983md,deBlas:2017xtg}. If this were true, a large $R^2$-coupling would imply equally large higher-order couplings, in which case the Starobinsky action would not be a good approximation at inflationary scales, and the truncation of the Lagrangian to the quadratic order would be unjustified.

Testing this expectation from first principles is the scope of our work, and requires a framework in which the RG flow of the gravitational couplings can be followed from a well-defined fundamental regime down to the energy scales relevant for inflation. Asymptotic safety (AS)~\cite{Weinberg:1980gg,Reuter:1996cp,Reuter:2012id,Eichhorn:2018yfc,Bonanno:2020bil} provides one such a framework. If the gravitational RG flow admits a UV fixed point with a finite number of relevant directions, then RG trajectories are well defined at all scales, and the values of the low-energy couplings---the EFT Wilson coefficients---are determined by the flow rather than set by hand. In particular, the number of free parameters in the EFT equals the number of relevant directions of the fixed point, minus one setting the overall unit scale. The functional renormalization group (FRG)~\cite{Dupuis:2020fhh} is the main non-perturbative tool to investigate this scenario. It has provided substantial evidence for the existence of a suitable fixed point, named the Reuter fixed point, including in truncations with higher-curvature operators and matter~\cite{Reuter:2001ag, Lauscher:2002sq, Codello:2006in,Machado:2007ea,Codello:2008vh,Benedetti:2009rx,Dona:2013qba,Falls:2013bv,Falls:2014tra, Gies:2015tca, Gies:2016con,Biemans:2016rvp,Biemans:2017zca, Christiansen:2017cxa, Hamada:2017rvn, Knorr:2017fus, Christiansen:2017bsy, Falls:2017lst,Knorr:2018fdu,Eichhorn:2019ybe, Alkofer:2020vtb,Falls:2020qhj, Knorr:2021lll, Knorr:2021slg, Kluth:2022vnq, Pastor-Gutierrez:2022nki,Baldazzi:2023pep,Eichhorn:2025sux}; this is a non-trivial feature, as AS is not a generic feature of perturbatively non-renormalizable theories~\cite{Heisenberg:2026rdk,DelPorro:2026wvw}. AS has also been used extensively in cosmology, including in the context of inflation~\cite{Bonanno:2001hi,Weinberg:2009wa,Bonanno:2015fga,Gubitosi:2018gsl}, showing that the nearly scale-invariant spectrum of primordial perturbations could be related to the approximate scale invariance of the RG flow near the Reuter fixed point. Remarkably, the Reuter fixed point comes with three relevant directions that, in an $f(R)$ setting, correspond to Newton's coupling, cosmological constant, and $R^2$-coupling~\cite{Reuter:1996cp,Benedetti:2013jk,Litim:2014uca,Falls:2014tra,Denz:2016qks,Eichhorn:2018yfc,Basile:2024oms,Eichhorn:2026uqj,Denz:2016qks}. As a consequence, the $R^2$-coupling is a free parameter in AS, and it can be set to match the cosmological data~\cite{Gubitosi:2018gsl}, particularly based on the amplitude of the scalar power spectrum. The question we address in this work is whether higher-derivative operators can spoil this result, based on how the RG generates and evolves couplings. Concretely, we use the FRG to study the flow of the couplings of a gravitational action including operators up to the fourth order, along all RG trajectories that depart from the UV fixed point at high energies. These are the UV-complete trajectories that define the AS landscape~\cite{Basile:2021krr,Knorr:2024yiu,Basile:2025zjc,DelPorro:2025wts,DelPorro:2025fiu}---the set of EFTs predicted by an AS UV completion. This allows us to make predictions for the relations among Wilson coefficients at inflationary scales\footnote{We note that this is a first-principles calculation that goes beyond the AS-inspired RG-improved setups~\cite{Dou:1997fg,Bonanno:2001hi,Bonanno:2001xi,Babic:2004ev,Reuter:2004nx,Reuter:2005kb,Bonanno:2005mt,Bonanno:2007wg,Bonanno:2008xp,Weinberg:2009wa,Tye:2010an,Koch:2010nn,Bonanno:2010mk,Bonanno:2010bt,Cai:2011kd,Capozziello:2011et,Hong:2011ws,Bonanno:2012jy,Hindmarsh:2012rc,Cai:2013caa,Copeland:2013vva,Xianyu:2014eba,Codello:2014sua,DOdorico:2015jtl,Bonanno:2015fga,Kofinas:2016lcz,Bonanno:2016rpx,Bonanno:2017gji,Bonanno:2017pkg,Bonanno:2018gck,Platania:2019qvo,Nagy:2025yam,Koch:2026xfv}}. 

Our original goal was to make the expectation described above quantitative, i.e., to show from a first-principles RG calculation that a large $R^2$-coupling accompanied by suppressed higher-order couplings is unnatural from an RG and EFT perspective. What we find instead is that the na\"ive expectation is to be refined, since only dimensionless Wilson coefficients associated with operators of the same mass dimension should be of the same order of magnitude. More precisely, in the fixed-point regime, the RG flow indeed generates dimensionless couplings of the same order of magnitude, regardless of their mass dimensions. However, once the flow leaves the fixed-point regime and canonical-like scaling takes over in the infrared (IR), the most irrelevant couplings are suppressed by the canonical scaling, i.e., by powers of the ratio between the running scale and the Planck scale. Combined with the large separation between the Planck scale and the scale of inflation, this produces strongly suppressed higher-order operators, which is exactly the hierarchy required for compatibility with cosmological data.

This result has two main implications. On the conceptual side, it refines the naturalness criterion for gravitational EFTs: what is natural is not that all dimensionless couplings are of the same order at every scale, but that couplings of the same mass dimension are, with the hierarchy between different operators controlled by the RG flow and by the separation of scales. On the phenomenological side, it shows that the Starobinsky action is not a fine-tuned truncation of the gravitational EFT, but is consistent with a gravitational EFT embedded in a UV-complete framework in which the $R^2$-coupling is a relevant direction. We highlight that, since higher-order operators are suppressed but not absent, they induce corrections to the Starobinsky predictions, which could be worth considering in light of the tension with the latest data~\cite{AtacamaCosmologyTelescope:2025blo,AtacamaCosmologyTelescope:2025nti}.
The paper is organised as follows. In Sect.~\ref{sect:setup} we set up the problem, introducing the gravitational EFT extending the Starobinsky action, collect the observational constraints on the higher-derivative Wilson coefficients, and explain how the RG connects the two. Sect.~\ref{sect:frg} reviews the FRG and its application to $f(R)$ gravity, fixing the flow equation we use throughout. Sect.~\ref{sect:fRweird} collects a number of features of $f(R)$ flows that are unusual compared with finite-dimensional truncations and that have a direct influence on our results. Readers interested only in the consequences of $f(R)$ flows for Starobinsky inflation may skip this section. Sect.~\ref{sect:unnaturalness} contains our main analysis and results. We first describe the patching strategy used to connect the fixed-point regime to inflationary scales, then determine the geometry of the AS landscape at sub-Planckian scales by integrating a large sample of UV-complete trajectories, and finally assess the naturalness of Starobinsky inflation from an FRG perspective. Sect.~\ref{sect:conclusions} summarises our findings and discusses their implications.

\section{Idea and setup: gravitational EFTs beyond Starobinsky}\label{sect:setup}

This section introduces the basic ingredients of our work: a gravitational EFT including the Starobinsky terms, the observational constraints on higher-derivative couplings, and our setup to connect the two via the RG. 

We shall consider a gravitational EFT which includes all terms allowed by symmetry. Limiting ourselves to scenarios relevant for inflation, the Lagrangian is given by a generic $f(R)$ theory. In Lorentzian signature, this reads
\begin{equation}
    f(R)=\frac{1}{16\pi G}\left(-2\Lambda+ R+b R^2+c R^3+ d R^4 +\dots \right)\,.
\end{equation}
In the context of Starobinsky inflation, the coefficient of the $R^2$-term is related to the inflaton mass $m$ via $b=1/(6m^2)$. This is constrained by the amplitude of the scalar power spectrum~\cite{Planck:2018jri}, yielding~\cite{Cheong:2020rao}
\begin{equation}
    \frac{1}{4\pi G (6m^2)}\simeq 2.26\cdot 10^9 \left( \frac{N_e}{60}\right)^2\quad \Rightarrow \quad m\simeq \left( \frac{60}{N_e}\right) 10^{13}\text{GeV}\,,
\end{equation}
with $N_e\sim50-60$ the number of e-folds during inflation. The bound on the amplitude constrains the $R^2$-coupling to be large, $b M_{\rm Pl}^2={M_{\rm Pl}^2}/{(6m^2)}=1/6 \cdot 10^{12}$. Inflation ought to occur at $k_{\rm infl}\sim 10^{13}-10^{16} \text{GeV}$. For definiteness, we shall consider $k_\text{infl}\sim 10^{13} \text{GeV}=10^{-6} M_{\rm Pl}$, where $M_{\rm Pl}=10^{19} \text{GeV}$ is the Planck mass. 

From the viewpoint of fundamental physics, the vanilla Starobinsky model---defined by a purely quadratic Lagrangian---is unnatural: any consistent (e.g., unitary) UV completion of gravity is expected to give rise to a tower of higher derivative operators. This could, in turn, spoil the agreement between the Starobinsky model and observational data: a pure $R^2$-action corresponds to an exactly scale-invariant theory, resulting in a flat scalar potential $V(\phi)$ in the Einstein frame. The Einstein-Hilbert term bends the potential downwards, creating the minimum needed for inflation to end and for reheating to happen; yet, higher derivative terms---unless suppressed---bend the potential upwards for $\phi\gg M_{\rm Pl}$, potentially preventing slow-roll inflation and spoiling agreement with observational data~\cite{Platania:2019qvo}.

On these grounds, compatibility of EFT corrections with the observational data has been studied extensively~\cite{Berkin:1990nu,Kaneda:2010ut,Ketov:2010qz,Huang:2013hsb,Miranda:2017juz,Cheong:2020rao,Ivanov:2021chn,Ketov:2025cqg,Addazi:2025qra, Drees:2025ngb}. As a paradigmatic example, considering a cubic extension to Starobinsky inflation, the Planck data constrain the cubic coupling to be parametrically suppressed~\cite{Cheong:2020rao},
\begin{equation}\label{eq:cheong-rel}
    \left|\frac{3c}{4 b^2}\right|\lesssim 10^{-4}\,.
\end{equation}
The simple model of~\cite{Cheong:2020rao} already shows that including even a single EFT correction beyond the Starobinsky model, as expected on theoretical grounds, tends to spoil agreement with observations unless the cubic Wilson coefficient is strongly suppressed relative to the quadratic one. Going beyond the cubic order, and embedding the Starobinsky Lagrangian in a gravitational EFT, compatibility with cosmological data yields a tower of constraints~\cite{Huang:2013hsb,Ketov:2025cqg}. Concretely, let us consider an $f(R)$ Lagrangian of the form~\cite{Huang:2013hsb,Ketov:2025cqg},
\begin{equation}\label{eq:action-ketov}
    f(R)=\frac{1}{16\pi G_\text{N}}\left(-2\Lambda+ R+\frac{c_2}{m^2} R^2+\frac{c_3}{m^4} R^3+ \frac{c_4}{m^6} R^4 +\dots \right)\,,
\end{equation}
where $m$ is the inflaton mass as defined above, $c_i$ are dimensionless Wilson coefficients, and 
\begin{equation}
     C_{i}=\frac{c_i}{m^{(2i-2)}}
\end{equation}
are the corresponding dimensionful couplings in units of the inflaton mass. Using the compatibility with data from the Atacama Cosmology Telescope (ACT)~\cite{AtacamaCosmologyTelescope:2025blo,AtacamaCosmologyTelescope:2025nti} and DESI~\cite{DESI:2024mwx}, one can constrain all coefficients $c_i$. For concreteness,~\cite{Ketov:2025cqg} found one specific set of values that reproduces the ACT/DESI central values $n_s = 0.974$ and $\alpha_s = +0.0062$ within a quintic model. These values are
\begin{equation}
    c_2=1/6\,,\quad c_3\simeq 3\cdot 10^{-4}\,,\quad c_4\simeq -2\cdot 10^{-6}\,,\quad c_5\simeq 4\cdot 10^{-9}\,, \label{eq:bounds}
\end{equation}
thereby confirming that higher-derivative corrections beyond the quadratic ones would generally spoil agreement with observations, unless the higher-derivative Wilson coefficients are increasingly suppressed relative to the quadratic one. Similar hierarchies have been found using different setups~\cite{Ivanov:2021chn,Bezerra-Sobrinho:2025gfg}\footnote{We note that some works find a positive $c_3$ to be in accordance with observations, e.g.~\cite{Ketov:2025cqg} while others seem to favor a negative sign, e.g.~\cite{Bezerra-Sobrinho:2025gfg}, especially after including ACT data releases. We will discuss how our results fit with these findings in Sect.~\ref{subsec:nat_full}, but we remark that we are mostly interested in the magnitude of the coefficients rather than the sign, since that is not firmly established.}. For general $f(R)$ actions of the form~\eqref{eq:action-ketov}, compatibility with data requires the following suppression for higher-derivative couplings~\cite{Huang:2013hsb}
\begin{equation}\label{eq:unn-val}
|c_n| \lesssim \frac{10^{-2n+2.6}}{2n\cdot3^{n-1}}\,.
\end{equation}
In this work, we will use an explicit RG computation to test whether the specific bound~\eqref{eq:cheong-rel} and the more general hierarchy~\eqref {eq:unn-val} are compatible with fundamental physics. The theoretical motivation is that in the regime where Quantum Field Theory (QFT) holds, the RG yields dimensionless Wilson coefficients $c_i$ of the same order~\cite{Manohar:1983md,deBlas:2017xtg}, in apparent tension with the constraints~\eqref{eq:unn-val}.

As it shall soon become clear, Eq.~\eqref{eq:action-ketov} is written in a convenient form for our purposes. RG computations are performed using a Euclidean signature, which amounts to an overall minus sign compared to the Lorentzian action above. Specifically, we shall use the FRG, which describes how the effective action changes with the RG scale $k$, and interpolates between the bare action in the UV, $k\to\infty$, and the full effective action in the IR limit, $k\to0$. To assess whether the hierarchy~\eqref{eq:unn-val} among the Wilson coefficients required by observations follows from the RG dynamics, we must first identify a suitable UV fixed point, which provides the starting point for the RG trajectory and allows us to make predictions for the low-energy coefficients in terms of the relevant directions of the fixed point. The FRG flow is naturally formulated in terms of dimensionless running couplings, which in our case are defined as follows 
\begin{equation}\label{eq:dimlessdimfull}
    \lambda_k=k^2 \Lambda_k\,,\quad g_k=k^{-2}G_k\,,\quad u_{i,k}=k^{2-2i}U_{i,k}\,.
\end{equation}
The scale-dependent effective action in Euclidean signature then reads
\begin{equation}
    \Gamma_k=\frac{k^2}{16\pi g_k}\int d^4 x\sqrt{g}\left(2\lambda_k k^2-R+\frac{u_{2,k}}{k^2} R^2+\frac{u_{3,k}}{k^4} R^3+ \frac{u_{4,k}}{k^6} R^4 +\dots\right)\,.
\end{equation}
Typically, in formal calculations, the dimensionless Wilson coefficients are identified as appropriate ratios of dimensionful running couplings in the physical limit $k\to0$. In simple systems in which there is a single scale, one may identify the RG scale $k$ with the physical momentum $p$ and still obtain qualitatively similar results (see~\cite{Pawlowski:2023gym} and references therein). We argue that the two strategies coincide in our case, for at least two reasons. First, we expect QG fluctuations to become suppressed below the QG scale, which in our case is $k_{\rm FP}= \sqrt{g_*} M_{\rm Pl}$, with $g_* \sim \mathcal{O}(1)$ based on numerous FRG calculations~\cite{Reuter:2001ag, Lauscher:2002sq, Codello:2006in,Machado:2007ea,Codello:2008vh,Benedetti:2009rx,Dona:2013qba,Falls:2013bv,Falls:2014tra, Gies:2015tca, Gies:2016con,Biemans:2016rvp,Biemans:2017zca, Christiansen:2017cxa, Hamada:2017rvn, Knorr:2017fus, Christiansen:2017bsy, Falls:2017lst,Knorr:2018fdu,Eichhorn:2019ybe, Alkofer:2020vtb,Falls:2020qhj, Knorr:2021lll, Knorr:2021slg, Kluth:2022vnq, Pastor-Gutierrez:2022nki,Baldazzi:2023pep,Eichhorn:2025sux}. Consequently, one can approximate $\Gamma_{k_{\rm infl}}\simeq \Gamma_0$. As an explicit test, using a well-established analytical approximation for the RG running of Newton's coupling~\cite{Bonanno:2000ep},
\begin{equation}\label{eq:g-run}
    G_k=\frac{G_{\rm N}}{1+g_*k^2 G_{\rm N}}\,,
\end{equation}
one can explicitly verify that at the inflationary scale $k_\text{infl}\simeq 10^{-6} M_{\rm Pl}$ the RG-scale-dependent Newton's coupling\footnote{Note that the Newton's coupling does not run with respect to physical momenta, but it depends on the RG scale $k$~\cite{Donoghue:2019clr,Bonanno:2020bil,Borissova:2022mgd,Basile:2024oms}. As $g_* \simeq \mathcal{O}(1)$~\cite{Reuter:2001ag, Lauscher:2002sq, Codello:2006in,Machado:2007ea,Codello:2008vh,Benedetti:2009rx,Dona:2013qba,Falls:2013bv,Falls:2014tra, Gies:2015tca, Gies:2016con,Biemans:2016rvp,Biemans:2017zca, Christiansen:2017cxa, Hamada:2017rvn, Knorr:2017fus, Christiansen:2017bsy, Falls:2017lst,Knorr:2018fdu,Eichhorn:2019ybe, Alkofer:2020vtb,Falls:2020qhj, Knorr:2021lll, Knorr:2021slg, Kluth:2022vnq, Pastor-Gutierrez:2022nki,Baldazzi:2023pep,Eichhorn:2025sux}, in the physical limit $k\to0$, the constant Newton's coupling $G_{\rm N}$ is recovered.} approximately yields the constant Newton's coupling that one expects in the IR limit,
\begin{equation}
    G_{k_\text{infl}}\simeq (1+10^{-12}g_*)^{-1} M_{\rm Pl}^{-2}\simeq G_{\rm N}\,.
\end{equation}
Note that the relation above for the running $G$ sets the transition scale to $k_\text{FP}=1/(g_* G_{\rm N})^{1/2}$.

Second is the decoupling mechanism. The RG flow can be halted whenever there is a characteristic decoupling scale $k_{\rm dec}$ in the IR, which inhibits the regulator in the Wetterich equation, thus freezing out the flow~\cite{Borissova:2022mgd}. In our case, the role of such an IR mass scale may be taken by the inflaton mass, $k_{\rm dec}\sim m$. 

These considerations are particularly useful in our case because, upon identifying the inflationary scale as $k_\text{infl}\simeq m$, we can directly read off the physical Wilson coefficients from the effective action
\begin{equation}
    \Gamma_0\simeq \Gamma_{k\simeq m}=\frac{1}{16\pi G_{\rm N}}\int d^4 x\sqrt{g}\left(2\Lambda-R+\frac{u_{2,m}}{m^2} R^2+\frac{u_{3,m}}{m^4} R^3+ \frac{u_{4,m}}{m^6} R^4 +\dots\right)\,,
\end{equation}
which is precisely of the form~\eqref{eq:action-ketov}, modulo a Wick rotation relating the Euclidean and Lorentzian coefficients, $c_i\equiv -u_{i,m}$. With this setup, we can compute the flow at $k\simeq m$ and directly compare the resulting physical Wick-rotated Wilson coefficients with the constraints~\eqref{eq:unn-val}. 

As a practical framework, we consider the RG flow in a QFT setting, where UV-complete trajectories are those emanating from the Reuter fixed point of asymptotically safe gravity. We expect the qualitative conclusion to hold also in other UV completions. Asymptotically safe quantum gravity provides a concrete setting in which this expectation can be tested through explicit calculations.

The Reuter fixed point comes with three relevant directions, corresponding to the couplings $g_k$, $\lambda_k$, $u_{2,k}$. These need to be fixed by experiments, while all other infinitely many couplings are irrelevant, meaning that they are predictions of the theory. Non-trivially, it has been shown~\cite{Gubitosi:2018gsl} that in a quadratic truncation of Asymptotically Safe Quantum Gravity (ASQG) there exist RG trajectories that are both UV complete and compatible with the observed values of $G_{\rm N}$, $\Lambda$, and $m^2$. In our work, we will make predictions for the couplings $c_{k\simeq m,i}$ and we will compare them with the values of $-c_i$ in Eq.~\eqref{eq:unn-val}\footnote{The minus sign here is due to the Wick rotation from Euclidean back to Lorentzian signature.}.

In the next section, we shall discuss in greater detail the FRG and its application to $f(R)$ theories.

\section{FRG for $f(R)$ gravity}
\label{sect:frg}

In this section, we briefly review the FRG approach and its development in $f(R)$ gravity, with the aim of summarizing the current status of RG flows in this setting and setting the stage for our analysis.

The FRG provides a non-perturbative framework to investigate the RG flow of a QFT. Its central object is the effective average action (EAA) $\Gamma_k$, which acts as an interpolant between the microscopic action, $\Gamma_{k\to\infty}= S$, and the full quantum effective action, $\Gamma_{k\to 0}=\Gamma$. Its evolution with the RG scale $t=\ln(k/k_0)$ is governed by the exact evolution equation~\cite{Wetterich:1992yh,Morris:1993qb}
\begin{equation}\label{eq:wetterich}
	k \partial_k \Gamma_k= \frac{1}{2}\mathrm{sTr}\left\{\left(\Gamma_k^{(2)}+\mathcal{R}_k\right)^{-1}k\partial_k \mathcal{R}_k\right\}\,.
\end{equation}
Here, $\Gamma_k^{(2)}$ denotes the Hessian of the EAA with respect to the fluctuation fields, $\mathrm{sTr}$ is the (super)trace over all internal indices and momenta, and $\mathcal{R}_k$ is an IR regulator implementing the Wilsonian shell-by-shell integration of modes by suppressing fluctuations with momenta below the scale~$k$. Therefore, the flow equation can be understood conceptually as describing the progressive integration of quantum fluctuations as the RG scale $k$ is lowered. Its non-perturbative character makes it particularly suitable for investigating the AS conjecture~\cite{Weinberg:1980gg}, i.e., the idea that gravity is UV-completed at an interacting fixed point of its RG flow. 

In practice, solving the Wetterich equation requires approximations, since the EAA should in principle span an infinite-dimensional theory space, so the flow cannot be solved exactly. A common strategy is therefore to truncate this to a manageable subset of operators, including functions of curvature invariants~\cite{Codello:2007bd, Machado:2007ea, Codello:2008vh, Benedetti:2012dx, Dietz:2012ic, Falls:2013bv,Benedetti:2013jk, Falls:2014tra, Ohta:2015efa, Falls:2018ylp,Kluth:2022vnq,Basile:2021euh,Basile:2021krk,Basile:2026ssn} or form factors~\cite{Knorr:2019atm,Knorr:2021iwv,Knorr:2022dsx}. In this respect, the FRG and $f(R)$ gravity are particularly well suited to one another, the former providing a systematic and non-perturbative framework for studying the RG flow of the latter, and the $f(R)$ ansatz retaining an infinite tower of curvature interactions within a single function rather than truncating at finite order in curvature. Specifically, for the purpose of our work, we allow the EAA to depend solely on the Ricci scalar,
\begin{equation}
\Gamma_k[g]=\int \mathrm{d}^4x\sqrt{g}f_k(R)\,,
\end{equation}
thus promoting the classical $f(R)$ function to be FRG scale-dependent, while neglecting independent invariants involving other contractions of the curvature tensor. Substituting this ansatz into the Wetterich equation and projecting the resulting flow onto its $f(R)$ sector yields a functional RG equation for $f_k(R)$. This equation is generally a nonlinear partial differential equation, where the scale dependence of the function is coupled to its derivatives with respect to curvature. 

FRG applications to $f(R)$ gravity have been developed extensively in the literature, primarily establishing and characterizing the UV structure of the theory by constructing the non-perturbative flows and exploring the fixed point structure~\cite{Codello:2007bd, Machado:2007ea, Codello:2008vh, Benedetti:2012dx, Dietz:2012ic, Falls:2013bv,Benedetti:2013jk, Falls:2014tra, Ohta:2015efa, Falls:2018ylp}. In particular,~\cite{Falls:2014tra}, subsequently extended in~\cite{Falls:2018ylp} with high-order recursive relations and resummation, examined the stability and convergence of the fixed-point solution with increasing order of the power series in $R$, providing further evidence for the existence of an interacting non-Gaussian fixed point (NGFP). In this work, we adopt the $f(R)$ flow equation derived in it, written in terms of the dimensionless curvature $r=k^{-2} R$, and for the dimensionless version of the function: $\hat f_k(r)=16 \pi f_k(R)/k^4$. The flow equation takes the form
\begin{equation}
    \partial_t \hat f + 4 \hat f -2 r \hat f'=\mathcal{I}[\hat f]\,, \label{eq:flow_eq_f}
\end{equation}
with
\begin{equation}
    \mathcal{I}[\hat f] = \mathcal{I}_0[\hat f] + \mathcal{I}_1[\hat f]\,\partial_t \hat f' + \mathcal{I}_2[\hat f]\,\partial_t \hat f''\,. \label{eq:I_split}
\end{equation}
Here, $\hat f'= \mathrm{d}\hat f/\mathrm{d}r$, and $\mathcal{I}[\hat f]$ encapsulates the contribution from quantum fluctuations in this specific setup; the explicit expressions for the functions $\mathcal{I}_{0,1,2}$ are given in the appendix of~\cite{Falls:2014tra}. We use this equation as the starting point for a polynomial expansion in $R$, allowing us to track the RG evolution of the corresponding curvature couplings as the truncation order is extended.

\section{Peculiar features of $f(R)$ flows}\label{sect:fRweird}

This section collects and discusses features of $f(R)$ flows that are unusual compared with those of other finite-dimensional truncations studied in AS. Some of these features have direct implications for our calculations and results. This discussion is primarily intended for readers interested in the technical details of gravitational RG flows; readers mainly interested in the consequences for Starobinsky inflation may skip it and proceed to the following section.

\subsection{Adaptive vs non-adaptive regulators and singularities in $f(R)$ flows}

Besides the choice of truncation, the flow equation depends on the regulator~$\mathcal R_k$  used in the Wetterich equation~\eqref{eq:wetterich}. For $f(R)$-like truncations, an important distinction is whether $\mathcal R_k$ depends on the function $f$ and its derivatives, in which case the regulator is called adaptive, or whether it is independent of them, in which case it is called non-adaptive~\cite{Benedetti:2012dx,Benedetti:2013jk,Mitchell:2021qjr,Morris:2022btf}. This choice affects the mathematical structure of the flow equation, and it is therefore relevant for the analytic control we have on the flow.

Adaptive regulators are introduced by replacing
\begin{equation}
    -\nabla^2 \;\mapsto\; -\nabla^2 + \mathcal R_k(-\nabla^2/k^2)
\end{equation}
directly in the Hessian $\Gamma^{(2)}_k$. This simplifies the evaluation of the traces and allows one to obtain the beta functions in an analytic form. 
The fixed-point structure stemming from these flows can be studied under the enlargement of the truncation, showing that the Reuter fixed point appears at each order~\cite{Litim:2001dt,Litim:2001fd,Litim:2002cf,Litim:2003vp,Fischer:2006at,Litim:2007jb}. 
However, the Hessian of the conformal mode contains $f''(R)$, so that $\partial_t \mathcal{R}_k$ produces a term proportional to $R\,f'''(R)$. The flow equation then becomes a third-order ordinary differential equation (ODE) in $R$, with a fixed singularity at $R=0$, where the coefficient of $f'''$ vanishes~\cite{Benedetti:2012dx,Morris:2022btf}. In general, further fixed singularities appear at $R>0$. Each of them imposes an additional condition on the solution, making the existence of global solutions difficult to establish~\cite{Morris:2022btf}.

Non-adaptive regulators, on the other hand, do not depend on $f$. The flow equation remains of second order in $R$ and is free of fixed singularities, so that fixed-point solutions are isolated and possess a finite number of relevant directions~\cite{Morris:2022btf}, as required for a predictive UV completion. The drawback is that the traces cannot be evaluated analytically, and the flow equation has to be solved numerically.
In this work, we employ the beta functions of~\cite{Falls:2014tra}, which were derived with an adaptive regulator. The reason behind our choice is that, as discussed above, in the case of adaptive regulators the beta functions can be obtained in a closed analytic form---which we need for our purposes. At the same time, as we shall see in the next sections, this makes the numerical integration of the flow quite involved.

\subsection{Relevant directions, IR shields, and enhanced predictivity}\label{subsect:shield}

For a given fixed point, the couplings associated with irrelevant directions are predictions of the specific UV completion, whereas the relevant couplings correspond to free parameters to be determined by experiment. This is often read as saying that a relevant coupling can take any value. In general it cannot. 

\begin{figure}[t!]
    \centering
    \includegraphics[width=0.7\linewidth]{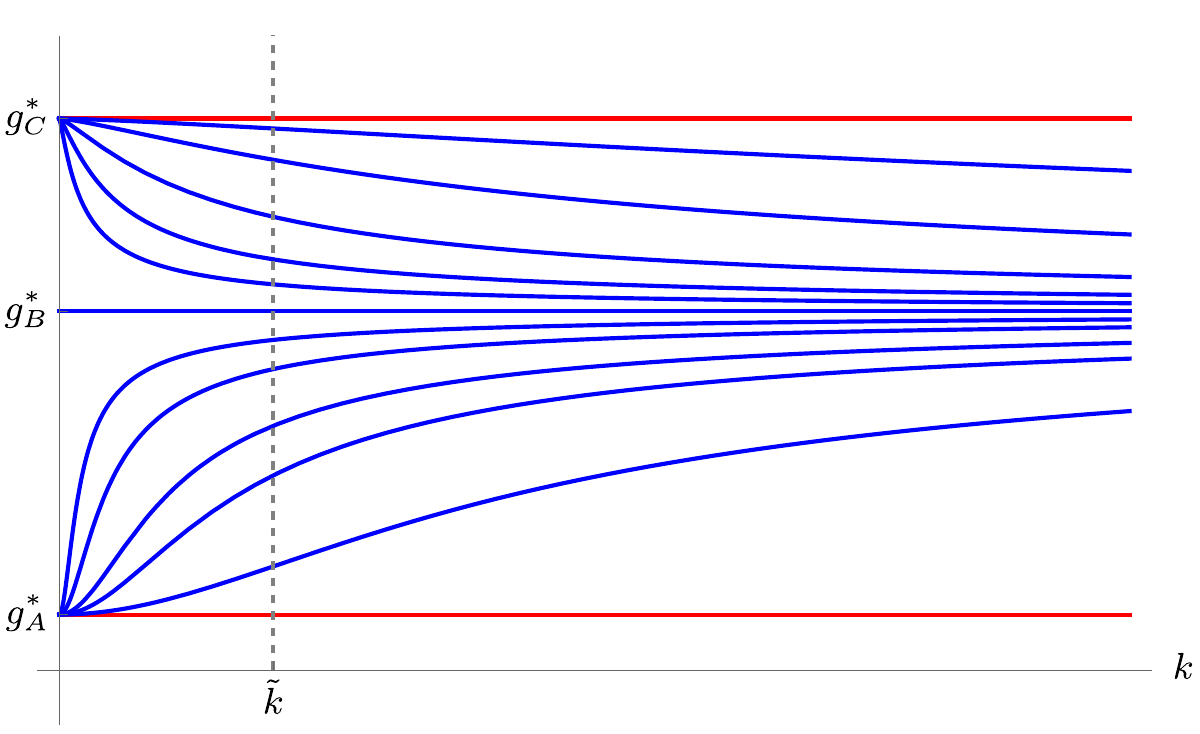}
    \caption{Effect of IR shields, from a toy example with three fixed points $\mathrm{FP}_A$, $\mathrm{FP}_B$
  and $\mathrm{FP}_C$. The central fixed point
  $\mathrm{FP}_B$ is UV attractive along $g$, so $g$ is a relevant direction
  there, while $\mathrm{FP}_A$ and $\mathrm{FP}_C$ are IR attractive along $g$. Trajectories emanating from $\mathrm{FP}_B$ run to $\mathrm{FP}_A$ or
  $\mathrm{FP}_C$ and cannot go beyond them. The two outer fixed points
  therefore act as IR shields, and the free parameter $g$ is restricted to
  $g^*_A < g < g^*_C$ at any finite $\tilde{k}\geq0$. This is key in our context, as due to IR shields in truncated $f(R)$ flows, the $R^2$-coupling---which is relevant in AS---cannot take any value at inflationary scales.}
    \label{fig:ir_shield}
\end{figure}

The rationale is that the global structure of an RG flow is shaped by all its fixed points, and any given system can, in principle, have more than one fixed point---each with its own combination of relevant and irrelevant directions. Concretely, let us consider a coupling $g$ that is relevant with respect to a fixed point $\mathrm{FP}_B$, and irrelevant with respect to two other fixed points $\mathrm{FP}_A$ and $\mathrm{FP}_C$. Flows leaving from the UV completion $\mathrm{FP}_B$ along $g$ are drawn towards either $\mathrm{FP}_A$ or $\mathrm{FP}_C$ in the IR, and trajectories cannot cross the values $g^*_A$ and $g^*_C$ along this direction. The allowed
range of $g$ is therefore bounded by $g^*_A$ and $g^*_C$ at any finite momentum (cf. Fig.~\ref{fig:ir_shield}). We call $\mathrm{FP}_A$ and $\mathrm{FP}_C$
\emph{IR shields} for the fixed point $\mathrm{FP}_B$,  in the direction $g$. At a practical level, in such situations, a relevant coupling is not entirely free, so that the net effect of IR shields\footnote{We would like to stress however, that in truncated versions of $f(R)$ theories, some of these IR shields may be truncation artifacts.} is an enhanced predictive power.

The $f(R)$ beta functions of~\cite{Falls:2014tra} provide an example of this mechanism, since there is more than one NGFP at any given truncation. While some of these may be truncation artifacts, they affect the whole RG flow at any given truncation. We shall see indeed in the next section that the range of values that $u_2$ can take is strongly limited by the presence of nearby IR shields. This fact will have an impact in Sect.~\ref{subsec:landscape}, in which the RG flow is run from the UV fixed point down to the scale $k\sim 10^{17}\text{GeV}$.

\subsection{IR features} \label{subsec:IR_features}

Since our aim is to investigate the implication of the RG running on Starobinsky inflation, we need to examine the IR structure of the $f(R)$ RG flow. We begin by discussing the appearance, or lack of, of the Gaussian fixed point (GFP) in different parametrizations of the EAA, and then examine the structure of the linearized flow around it. The main takeaway is that the IR structure is more involved than canonical scaling alone would suggest: even around the GFP, the linearized flow exhibits non-trivial mixing between couplings and logarithmic scale dependence. We then discuss the IR running about $k=0$, clarifying the role of the cosmological constant as $k\to0$.

\subsubsection{Placement of $G_{{N}}$ and existence of the GFP}

In this section we discuss the existence of the GFP, i.e., the fixed point characterized by vanishing interaction couplings, sitting at the origin of theory space. Analyzing the beta functions of~\cite{Falls:2014tra}, we found that its appearance in the RG flow depends on how Newton's coupling enters the EAA. Consider first an expansion of the form
\begin{equation}
    \Gamma_k=\int d^4 x\sqrt{g}\left(\frac{2\lambda_kk^4-k^2 R}{16\pi g_k}+{v_{2,k}} R^2+\frac{v_{3,k}}{k^2} R^3+ \frac{v_{4,k}}{k^4} R^4 +\dots\right)\,,\label{eq:no_overall_GN_EAA}
\end{equation}
with $v_{i,k}\equiv u_{i,k}/g_k$. Inserting this ansatz into the $f(R)$ flow equation, Eq.~\eqref{eq:flow_eq_f}, and expressing the flow in terms of the dimensionless couplings above, $(\lambda_k,g_k,v_{i,k})$, one finds no GFP. This feature is absent when instead all curvature terms are assigned the same overall Newton-coupling prefactor,
\begin{equation}
    \Gamma_k=\frac{k^2}{16\pi g_k}\int d^4 x\sqrt{g}\left(2\lambda_k k^2-R+\frac{u_{2,k}}{k^2} R^2+\frac{u_{3,k}}{k^4} R^3+ \frac{u_{4,k}}{k^6} R^4 +\dots\right)\,. \label{eq:overall_GN_EAA}
\end{equation}
In this parametrization, the origin of theory space is a fixed point of the RG flow, and the canonical mass dimensions of the couplings are reflected in the eigenvalues of the linearized flow---with caveats that will be discussed in the following subsections. The difference between these two descriptions can be traced to the transformation relating their coupling coordinates: since, schematically, the dimensionful couplings satisfy $v_{i}=u_{i}/g$, the transformation is singular at $g=0$, and its Jacobian is not invertible there. Consequently, the two parametrizations are not related by a regular coordinate transformation at the GFP: a GFP at the origin in the second parametrization need not map to a fixed point in the first. In particular, the absence of the GFP in the parametrization of Eq.~\eqref{eq:no_overall_GN_EAA} should be understood as an artifact of this singular coordinate transformation. 

\subsubsection{IR running about the GFP: non-canonical scalings and logarithmic form factors}

Even in the parametrization of Eq.~\eqref{eq:overall_GN_EAA} where the GFP exists, the linearized flow does not display the expected canonical scaling. In fact, for the truncation retaining $(\lambda,g,u_2,u_3)$, the stability matrix at the GFP reads
\begin{equation}
M=
\begin{pmatrix}
-2 & \alpha &  0 & 0\\
0 & 2 & 0 & 0 \\
0 & \beta & 2 & 0 \\
0 & \gamma & 0 & 4 
\end{pmatrix}\,,
\end{equation}
where $\alpha$, $\beta$, and $\gamma$ are constants determined by the flow equation. Its eigenvalues are $(-2,2,2,4)$, in agreement with the canonical scaling. However, the eigenvalue $2$ has algebraic multiplicity two and geometric multiplicity one. Therefore, the stability matrix is not diagonalizable: the eigenvectors do not form a complete basis, and it contains a non-trivial Jordan block.

To expose this structure, we introduce the following linear combinations of couplings
\begin{equation}
    w_0=\lambda-\frac{\alpha}{4} g\,, \qquad w_1=u_2\,, \qquad
    w_2=\beta g\,, \qquad w_3 = u_3+\frac{\gamma}{2} g \,,
\end{equation}
reducing the stability matrix to Jordan form. Here and below, the $k$ subscript on the couplings is suppressed for notational simplicity. The linearized beta functions are coupled,
\begin{equation}
    \partial_t w_0=-2 w_0\,, \qquad \partial_t w_1=2 w_1+w_2\,, \qquad \partial_t w_2=2 w_2\,, \qquad \partial_t w_3=4 w_3 \, ,
\end{equation}
and the non-trivial Jordan block in the $(w_1,w_2)$ sector generates a logarithm already at the linearized level. Moreover, translating back to the coupling basis of Eq.~\eqref{eq:overall_GN_EAA}, their IR running is now explicitly coupled
\begin{align}
\lambda(k)&=
\left(\lambda_0-\frac{g_0 \alpha}{4}\right)
\left(\frac{k}{k_0}\right)^{-2}
+\frac{g_0 \alpha}{4}\left(\frac{k}{k_0}\right)^2\,, \qquad
g(k)=g_0\left(\frac{k}{k_0}\right)^2\,,\label{eq:runninglgCC0} \\
u_2(k)&=\left( \frac{k}{k_0} \right )^2 \left [u_{2,0}+\beta g_0 \log\left(\frac{k}{k_0} \right) \right ]\,,\label{eq:u2_log}\\
u_3(k)&=u_{3,0}\left(\frac{k}{k_0}\right)^4
+\frac{\gamma}{2}g_0
\left[
\left(\frac{k}{k_0}\right)^4
-\left(\frac{k}{k_0}\right)^2
\right]\,.\label{eq:runningu3GFP}
\end{align}
Thus, the couplings $(\lambda,g,u_2,u_3)$ do not follow the canonical scaling one would expect for them in the IR, not only considering the logarithmic running. Indeed, for non-vanishing $g_0$, the $k^2$ contribution dominates the IR behavior of $u_3$, rather than its canonical $k^4$ contribution. This is a proper feature of the structure of the linearized flow.

Finally, let us consider the truncation including the $R^4$ operator. This adds an eigenvalue $6$, in agreement with the mass dimension of $u_4$, and a new mixing term in the $g$-column, while leaving the Jordan block in the $(g,u_2)$ sector unchanged. The Jordan form is obtained by extending the previous change of coordinates to include
\begin{equation}
    w_4=u_4+\frac{\delta}{4}g \,.
\end{equation}
The flow of the previously considered couplings, including the logarithmic running of $u_2$, remains unchanged. The additional coupling $u_4$ also exhibits mixing when expressed in the original coupling coordinates, meaning that its behavior about the GFP is not governed solely by its canonical scaling
\begin{equation}
    u_4(k)= u_{4,0} \left(\frac{k}{k_0}\right)^6 +\frac{\delta}{4}g_0\left[
    \left(\frac{k}{k_0}\right)^6-\left(\frac{k}{k_0}\right)^2 \right]\,.\label{eq:runningU4CC0}
\end{equation}
As for $u_3$, the induced $k^2$ contribution dominates as $k\to0$. 

More generally, the non-alignment of the eigendirections with the coordinate axes is not peculiar to $f(R)$ gravity, and it has also been observed in other settings~\cite{Codello:2008vh}. Here we go beyond the identification of this feature by investigating how this modifies the IR scaling of couplings---a crucial ingredient to extract the predictions we need for the Wilson coefficients. 

\subsection{IR running about $k=0$: role of the cosmological constant}\label{subsec:IR_beyond}

We showed that the linearized flow around the GFP already exhibits logarithmic corrections to its expected canonical scalings, as well as mixing between them. However, these results describe the flow in a neighborhood of the GFP and need not describe its behavior as $k \to 0$. This distinction is particularly important in the presence of the cosmological constant, which behaves like a mass scale: since its dimensionful and dimensionless counterparts are related by $\Lambda(k)=k^2\lambda(k)$, for any trajectory reaching a finite but non-vanishing $\Lambda_0\equiv \Lambda(k=0)$, one must have $\lambda(k)\simeq\Lambda_0\,k^{-2}$ in the IR, which diverges as $k\to0$. An expansion around small dimensionless couplings is therefore not, in general, equivalent to an expansion about $k=0$. Small-coupling limit and vanishing FRG scale, $k \to 0$, describe the same regime only if the observed cosmological constant is \emph{exactly} zero. The IR behavior must accordingly be investigated using an expansion adapted to the specific RG trajectory: around $\lambda=0$ if the dimensionful cosmological constant $\Lambda_0$ vanishes exactly, and around $\lambda=\infty$ otherwise.

First, let us consider the case in which $\Lambda_0=0$. This case corresponds to the unique trajectory connecting the NGFP in the UV with the GFP in the IR. For this so-called \emph{separatrix} line, the expansion about the GFP consistently describes the IR regime. Expanding the beta functions around small couplings and imposing $\lambda(0)=0$, we obtain
\begin{align}
\lambda(k)&=
\frac{g_0 \alpha}{4}\left(\frac{k}{k_0}\right)^2\,, \qquad g(k)=g_0\left(\frac{k}{k_0}\right)^2\,,\label{eq:runningR3CC0lg}\\
u_2(k)&=\left( \frac{k}{k_0} \right )^2 \left [u_{2,0}+\beta g_0 \log\left(\frac{k}{k_0} \right) \right ]\,,\\
u_3(k)&=u_{3,0}\left(\frac{k}{k_0}\right)^4
+\frac{\gamma}{2}g_0
\left[
\left(\frac{k}{k_0}\right)^4
-\left(\frac{k}{k_0}\right)^2
\right]\,. \label{eq:runningU3CC0}
\end{align}
This is exactly the linearized solution of
Eq.~\eqref{eq:u2_log} restricted to the separatrix line: the boundary condition $\lambda(0)=0$ removes the $k^{-2}$ scaling of the cosmological constant, which is then entirely induced by its mixing with Newton's coupling, while the Jordan-block logarithm in $u_2$ and the mixing-induced $k^2$ term in $u_3$ survive down to $k\to0$. On the separatrix, the expansion about the GFP and the expansion about $k=0$ therefore coincide. In what follows, we use this solution as the reference against which the case $\Lambda_0\neq0$ is compared.

For a non-vanishing dimensionful cosmological constant, $\Lambda_0\neq0$, instead, the beta functions must be expanded around large $\lambda$. The IR scaling reads
\begin{align}
\lambda(k)&=\lambda_0\left(\frac{k}{k_0}\right)^{-2}\,, \qquad
g(k)=g_0\left(\frac{k}{k_0}\right)^{2}\,,\\
u_2(k)&=\left(\frac{k}{k_0}\right)^{2}\left[u_{2,0}+\hat\beta\, g_0\log\left(\frac{k}{k_0} \right)\right]\,,\\
u_3(k)&=u_{3,0}\left(\frac{k}{k_0}\right)^{4}+\hat\gamma\, g_0\left(\frac{k}{k_0}\right)^{2}\,,
\end{align}
where the values of $(\hat\beta, \, \hat \gamma)$ differ from the linearized-flow counterparts: the functional form of the scalings coincides with that of the separatrix case, but the coefficients do not. The IR scalings are therefore discontinuous in $\Lambda_0$: the limits $k\to0$ and $\Lambda_0\to0$ do not commute, and a cosmological constant, however small, changes the IR behavior of the flow.

Finally, we extend the analysis to include the $R^4$ operator. On the separatrix, $\Lambda_0=0$, the picture is unchanged: the expansion around $\lambda=0$ reproduces the linearized solution for $u_4$ found in the previous section, with the canonical $k^6$ scaling supplemented only by the mixing-induced $k^2$ term. The solutions for $\lambda$, $g$, $u_2$ and $u_3$ are untouched. The case $\Lambda_0\neq0$ is instead affected dramatically. Expanding around $\lambda=\infty$, we obtain
\begin{align}
\lambda(k)&=\lambda_0\left(\frac{k}{k_0}\right)^{-2}\,, \qquad
g(k)=g_0\left(\frac{k}{k_0}\right)^{2}\,,\\
u_2(k)&=\left(\frac{k}{k_0}\right)^{2}\left[u_{2,0}+\hat\beta\, g_0\log\left(\frac{k}{k_0} \right)\right]\,,\\
u_3(k)&=u_{3,0}\left(\frac{k}{k_0}\right)^{4}+\hat\gamma\, g_0\left(\frac{k}{k_0}\right)^{2}\,,\\
u_4(k)&= \left(\frac{k}{k_0}\right)^6 \left [ u_{4,0} + \hat \epsilon \frac{g_0^2}{\lambda_0} \log\left(\frac{k}{k_0} \right) \right] +\frac{\hat \delta}{4}g_0\left[
    \left(\frac{k}{k_0}\right)^6-\left(\frac{k}{k_0}\right)^2 \right]\,.
\end{align}
With the quartic coupling, the $\Lambda_0\neq0$ solution differs from the separatrix one by more than the values of the coefficients, and, besides the mixing-induced $k^2$ term, $u_4$ acquires a logarithmic correction to its canonical $k^6$ scaling. This has no counterpart in the expansion around the GFP, although such logarithmic corrections may be suppressed by the fluctuations of massive matter fields~\cite{Knorr:2026vax}.

Strictly speaking, the true IR asymptotics of any trajectory with $\Lambda_0\neq0$ is captured by the expansion around $\lambda=\infty$, which includes the logarithmic correction to $u_4$. Such a structure, if still valid at inflationary scales, would have direct implications for the (un)naturalness of Starobinsky inflation. The deep IR is, however, not the relevant regime for our analysis. The expansion around $\lambda=\infty$ only holds when $k^2\ll \Lambda_0 $. Since, in addition, in our Universe $\Lambda_0\sim10^{-122}M_{\rm Pl}^2$, the observationally interesting trajectory must lie very close to the separatrix at inflationary scales. On these grounds, for the purpose of extracting the Wilson coefficients at $k\simeq m$, the GFP expansion provides the meaningful description, and it is the one we adopt in the remainder of this work.

\section{(Un)naturalness of Starobinsky inflation}
\label{sect:unnaturalness}
We now turn to the question of whether Starobinsky inflation can be regarded as a(n) (un)natural outcome of the RG evolution of higher derivative couplings. 

To assess the naturalness of Starobinsky inflation, we must account for several interconnected ingredients. We add them one at a time, which lets us disentangle which of them favor the required hierarchy of couplings and which disfavor it. Each step makes the analysis more involved, but it also brings us closer to a self-consistent picture. We reach our final assessment by including all ingredients simultaneously in the last subsection.

\subsection{Running from the UV down to inflationary scales: patching strategy} 

Ideally, one would determine the inflationary predictions by integrating the full system of beta functions, including all couplings up to fourth order in curvature, from the AS fixed point down to the inflationary scale. In practice, however, this approach is technically challenging due to the complexity of the coupled beta functions and the numerical difficulties involved in integrating the full system over the required range of scales. To overcome these difficulties, we adopt a \textit{patching strategy}, in which the RG evolution from the UV fixed point is determined in two steps: 
\begin{enumerate}
    \item \textit{Flow from the UV to sub-Planckian scales}: First, we integrate the beta functions as far into the sub-Planckian regime as possible, and provide as an output specific relations between the irrelevant and relevant couplings.
    \item \textit{Flow from sub-Planckian to inflationary scales:} Second, we use the derived relations at sub-Planckian scales as inputs for the subsequent IR evolution of couplings, which is dictated by how they scale about the GFP. 
\end{enumerate}
 
One can assess the qualitative validity of the patching procedure by examining the running of $g_k$. At low energies, quantum gravitational fluctuations are suppressed, and Newton's coupling approaches its constant, observed value $G_{\rm N}\simeq \text{const.}\equiv g_k/k^2$. One therefore expects $g_k$ to scale canonically, so that
\begin{equation}
    g_{k_{\rm infl}}/k_{\rm infl}^2\simeq g_{k_{\rm FP}}/k_{\rm FP}^2\,.
\end{equation}
This expectation can be tested against the full RG running given in Eq.~\eqref{eq:g-run}, which determines the scale dependence of the dimensionless Newton coupling,
\begin{equation}
    g_k=\frac{G_{\rm N} k^2}{1+(k/k_{\rm FP})^2}\,.
\end{equation}
It follows that
\begin{equation}
    \frac{G_{\rm N} k_{\rm infl}^2}{1+(k_{\rm infl}/k_{\rm FP})^2}k_{\rm infl}^{-2}\simeq\frac{G_{\rm N} k_{\rm FP}^2}{1+(k_{\rm FP}/k_{\rm FP})^2}k_{\rm FP}^{-2}\,.
\end{equation}
This holds provided we approximate $1/(1+10^{-12})\sim 1/2$, which is an order-one estimate and therefore sufficient for our purposes. In particular, based on the above relation, canonical scaling provides an excellent approximation for $k\in[0,10^{-1}M_{\rm Pl}]$. 
We conclude that, within a $\mathcal{O}(1)$-error, the canonical scaling is a good approximation below $k_{\rm FP}\simeq M_{\rm Pl}$, with the accuracy improving as $k$ decreases. Overall, the patching procedure provides a tractable approximation to the full RG evolution, especially if the matching is done at a sufficiently sub-Planckian scale $\bar{k}$.

As this will soon become important, we remark that the NGFP has three relevant directions, associated with the cosmological constant, Newton's, and $R^2$ couplings; the RG flow then predicts all other higher-derivative couplings, corresponding to the irrelevant directions of the Reuter fixed point, as a function of the relevant ones. The latter are free parameters, and at inflationary scales, $k_{\rm infl}\simeq 10^{-6}M_{\rm Pl}$, they can be chosen to match observations. Based on the definitions~\eqref{eq:dimlessdimfull}, this requires
\begin{equation}\label{eq:obs-values}
    \lambda(k_{\rm infl}) = 10^{-102}\,,\quad g(k_{\rm infl}) = 10^{-12} \,,\quad u_2(k_{\rm infl}) = -1/6\,.
\end{equation}
Finally, we report here the standard canonical scaling of dimensionless couplings, which follows from the definitions~\eqref{eq:dimlessdimfull}, requiring that in the IR each dimensionful coupling is approximately constant:
\begin{equation}
  \begin{aligned}
    \lambda(k) &= \lambda(k_{\rm infl})\,\left(\frac{k}{k_\text{infl}}\right)^{-2}\,, \quad g(k) = g(k_{\rm infl})\,\left(\frac{k}{k_\text{infl}}\right)^{2}\,,\quad
    u_{i\geq 2}({k}) &= u_i(k_{\rm infl}) \left(\frac{k}{k_\text{infl}}\right)^{2i-2}\,, 
  \end{aligned}
  \label{eq:canonicalscalingsimple}
\end{equation}
where, for convenience, the inflationary scale has been chosen as a reference scale. In the next subsection we will use simple approximations for the first part of the patching strategy and the canonical scaling above to get a first intuitive understanding of what can influence the naturalness question.

\subsection{Getting a grasp: importance of the geometry of the AS landscape} \label{FP&geom}

In this section we investigate the relations between irrelevant and relevant couplings using different approximations for the first part of the patching, and connecting it to a standard canonical scaling for the second part. While the approximations of this section are not ideal and will be relaxed later, this first approach will highlight, through semi-analytical approximations, the important role played by the geometry of the AS landscape in determining the naturalness of Starobinsky inflation.

\paragraph{Fixed-point approximation for $(\lambda_k,g_k)$: single RG trajectory.} \label{sec:FPapprox}

We begin with a simplified approximation in which the Newton's coupling and the cosmological constant are held fixed at their fixed-point values, $g_k\simeq g_*$ and $\lambda_k\simeq \lambda_*$. This first case provides a useful toy model to pinpoint the role of the AS landscape and to show, via simple analytical relations, how to connect AS predictions to observations at inflationary scales.

\begin{figure}[t!]
  \hspace{-0.7cm}\includegraphics[width=0.49\textwidth]{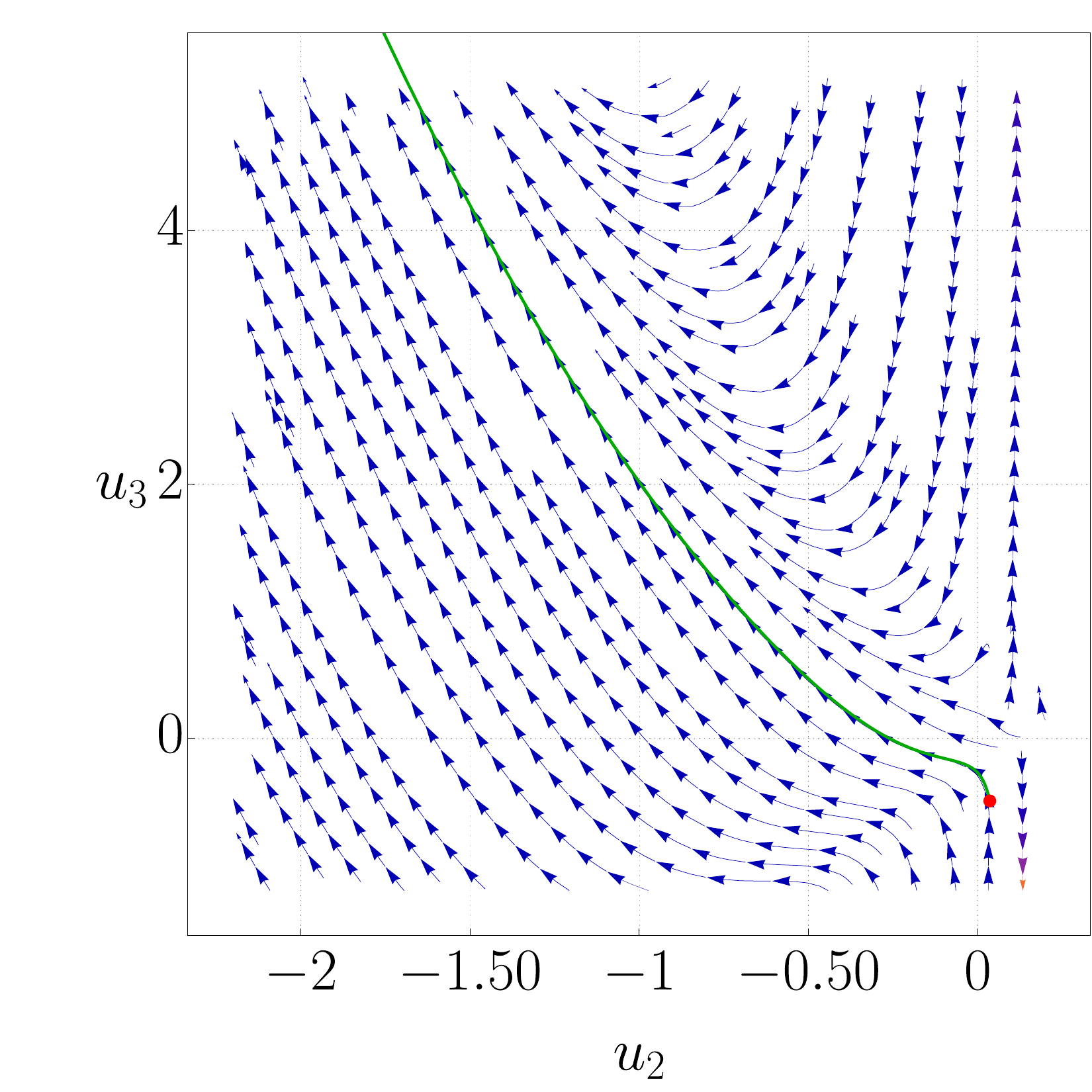}\qquad
  \includegraphics[width=0.49\textwidth]{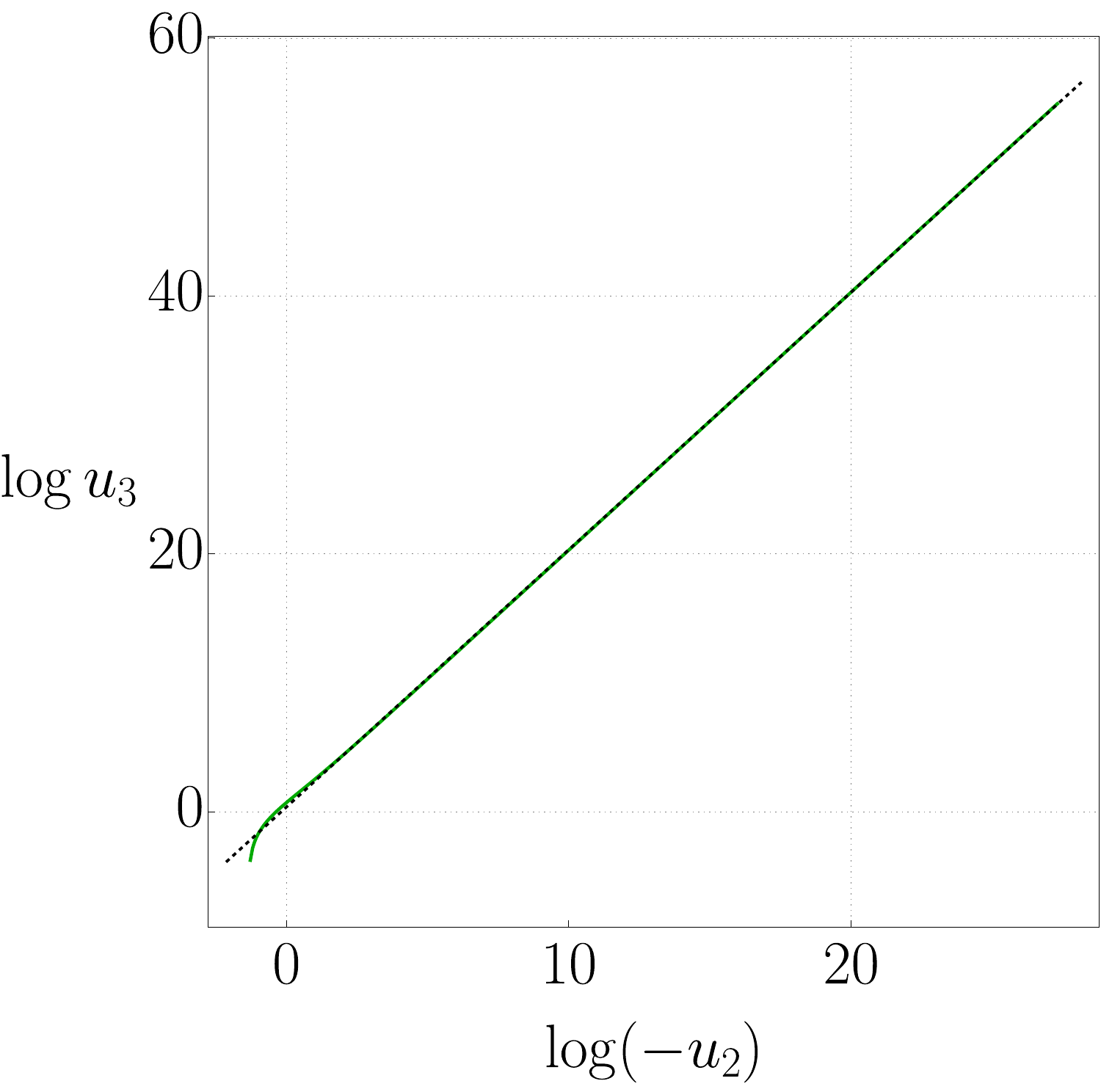}
  \caption{The left panel shows the flow in the $(u_2,u_3)$ plane at fixed $\lambda_*$ and $g_*$, with arrows pointing towards the IR. The red dot is the NGFP and the green curve is the trajectory leaving it. The right panel depicts the same trajectory in a log-log scale, as well as the best-fit trajectory (dashed line) of Eq.~\eqref{eq:loglogrel}.}
 \label{fig:fpapprox}  
  \end{figure}

Within this fixed-point approximation, the system is left with one relevant ($u_2$) and one irrelevant ($u_3$) coupling. In this setup, there is a single UV-complete trajectory.
In the case $\mathcal{I}_1=\mathcal{I}_2=0$ (cf.~Eq.~\eqref{eq:I_split}), the RG flow is particularly simple, and is shown in Fig.~\ref{fig:fpapprox}. Along the unique safe trajectory, the cubic and quadratic curvature couplings satisfy the relation
\begin{equation}\label{eq:loglogrel}
    \frac{u_3(k)}{u_2^2(k)}\simeq 1.58\,.
\end{equation}
Assuming this relation holds down to the scale $k_\text{FP}$, it can be combined with the canonical scaling~\eqref{eq:canonicalscalingsimple} and the observational values~\eqref{eq:obs-values}, to relate couplings at Planckian and inflationary scales, to infer that
\begin{equation}
    1.58=\left|\frac{u_{3,k_{\rm FP}}m^4/k_{\rm FP}^4}{ u_{2,k_{\rm FP}}^2(m^2/k_{\rm FP}^2)^2}\right| \simeq \left|\frac{u_{3,m}}{ u_{2,m}^2}\right| \nless \frac{4}{3}\cdot 10^{-4}\,.
    \label{eq:ratio_R3}
\end{equation}
Namely, within this first, rough approximation, the hierarchy required by data, Eq.~\eqref{eq:cheong-rel}, appears unnatural from a fundamental physics perspective. 

This result comes, however, with important caveats: \emph{(i)} it relies on the fixed-point approximation for $g_k$ and $\lambda_k$, \emph{(ii)} it sets $\mathcal{I}_1=\mathcal{I}_2=0$, and \emph{(iii)} it provides a fit to the RG trajectory itself, rather than the values that $u_3(k)$ can take as a function of $u_2(k)$ at a given sub-Planckian $k$. Based on the discussion in the previous section, the first of these approximations is not expected to provide an accurate description of the RG flow over the entire range of scales relevant to our analysis. Indeed, extrapolating the fixed-point approximation $G_k k^2= g_k\simeq g_*$ to inflationary scales would yield
\begin{equation} 
\label{eq:FPapprox_validation}
    G_m^{\rm FP}\sim g_* m^{-2}=10^{12}g_*G_{\rm N}\,.
\end{equation}
This is not accurate, unless $g_*\ll1$, which would amount to lowering the fixed-point regime to inflationary scales. We therefore relax these assumptions in turn. We first let $g_k$ and $\lambda_k$ run, turn the $\mathcal{I}_i$ on, and extend the analysis to include the $R^4$-coupling. Instead of using a single RG trajectory, we shall use an expansion about the fixed point to obtain general relations among all running couplings. Only in the next subsection, we will extend the analysis to a full-fledged numerical sampling a large number of UV-complete trajectories. 

\paragraph{Running $(\lambda_k,g_k)$: towards the AS landscape.}
\label{sec:tangent-plane}

The quadratic relation~\eqref{eq:loglogrel} between the $u_i$ couplings relies on the fixed-point approximation for $(\lambda_k,g_k)$ and on having set $\mathcal{I}_1=\mathcal{I}_2=0$. Relaxing both these approximations already introduces some challenges since they make the flow more involved, and there is no longer a single UV-complete trajectory. Hence, as an intermediate step, in this subsection we shall allow all couplings to run, and we will consider the full set of UV-complete trajectories, but we will focus on a linearized flow around the UV fixed point. For $\mathcal{I}_1=\mathcal{I}_2=0$, the linearized flow reads
\begin{equation}
u_3 \simeq -1.209 - 11.974\, u_2 + 0.026\, g +
8.450\,\lambda\,,\qquad \text{for }k \gg M_\text{Pl}\,.
\label{eq:tangent_cubic_a}
\end{equation}
This equation relates the value of the only irrelevant coupling in the system to the relevant ones, and describes the hypersurface of couplings at some $k \gg M_\text{Pl}$, which are UV-completed at the AS fixed point as $k\to\infty$. 
If we assume for a moment that this relation holds up to a sub-Planckian scale $\bar{k}$, one could then use it in combination with the canonical scaling~\eqref{eq:canonicalscalingsimple} to map the irrelevant coupling $u_3(\bar{k})$ to its value at the inflationary scale $k_{\rm infl}$. This would give $u_3(k_{\rm infl})\simeq
2.0\times10^{-12}$, well within the bound~\eqref{eq:bounds} and in sharp contrast with the result in Eq.~\eqref{eq:ratio_R3}. Including $\mathcal{I}_1$ and $\mathcal{I}_2$, which enter through the non-adaptive nature of the regulator used in~\cite{Falls:2014tra}, the linear relation above becomes
\begin{equation}\label{eq:tangent_cubic_b}
u_3 \simeq -0.501 + 1.169\, u_2 - 0.070\, g + 0.269\,\lambda\,,\qquad \text{for }k \gg M_\text{Pl}\,.
\end{equation}
This yields $u_3(k_{\rm infl})\simeq-1.9\times10^{-13}$, consistent with the $\mathcal{I}_1=\mathcal{I}_2=0$ result, and again in agreement with the observational bound~\eqref{eq:unn-val}.

Extending this procedure by including the $R^4$-coupling also yields
\begin{align}
\mathcal{I}_1=\mathcal{I}_2=0&:\quad u_3(k_{\rm infl})\simeq5.48\times10^{-13}\,,\quad
u_4(k_{\rm infl})\simeq-1.21\times10^{-24}\,, \label{eq:tangent_quartic_a}\\
\mathcal{I}_1,\mathcal{I}_2\neq0&:\quad u_3(k_{\rm infl})\simeq-1.88\times10^{-13}\,,\quad
u_4(k_{\rm infl})\simeq-2.10\times10^{-25}\,,
\label{eq:tangent_quartic_b}
\end{align} 
in agreement with the observational constraints~\eqref{eq:cheong-rel} and~\eqref{eq:unn-val}. It therefore appears that a proper suppression of the higher-order couplings can occur if these are related linearly to the relevant interactions, whereas a quadratic relation would lead to an incompatibility with observations. In spite of the approximations we have used, the considerations here highlight that \textit{the geometry of the AS landscape plays an essential role}. Specifically, in this subsection, we have accounted for the $\mathcal{I}_i$ corrections and for the running of all couplings, but we have derived the relations between relevant and irrelevant couplings through an expansion about the UV fixed point, which makes the linear relations between the couplings obvious. A crucial question is then whether evolving the UV-complete trajectories down to sub-Planckian scales would preserve the linearity of such relations among couplings, or whether it would bend the hypersurface describing their endpoints. We examine this question in the next subsection, while the interplay between geometry of the AS landscape, canonical scaling, and naturalness will be discussed, focusing on more conceptual aspects, in Sect.~\ref{subsec:nat_canonical}.

\subsection{Sampling all UV-complete trajectories: AS landscape at sub-Planckian scales}\label{subsec:landscape}

In this section we perform a full-fledged integration of the RG flow, from the UV fixed point down to sub-Planckian scales. This extends the results of the previous subsection in two ways: first, we no longer rely on approximations for the running couplings or for the flow; second, we do not study a single RG trajectory, but rather all UV-complete trajectories. Besides being more consistent, this strategy allows us to properly discuss (un)naturalness as a feature of the RG flow rather than of specific trajectories. The ultimate goal of this subsection is to obtain an equation relating irrelevant and relevant higher-derivative couplings at a given sub-Planckian scale, as predicted by an AS UV completion.

Following the strategy of~\cite{Basile:2021krr,Knorr:2024yiu}, we identify the AS landscape with the set of IR endpoints of RG trajectories emanating from the UV fixed point. We solve the full flow from the UV for a large number of initial conditions uniformly spanning the UV critical surface and covering all three relevant directions, with the goal of getting as close as possible to the IR scale of inflation. We implement this for the $f(R)$ truncation both up to $R^3$ and $R^4$, where the curvature couplings are rescaled by the Newton coupling, so that the GFP is present, as discussed in Sect. \ref{subsec:IR_features}. A crucial difference from the original procedure in~\cite{Basile:2021krr} is that we do not reach $k\rightarrow 0$. Every trajectory in our sample terminates at a finite scale, where either the beta functions become singular or the system becomes too stiff to integrate further, so we are only able to construct a landscape at a sub-Planckian scale common to all trajectories, rather than a proper IR one. This is sufficient for our purposes because it is possible to extract a relation between couplings to fix the irrelevant couplings at the transition scale $k_{\rm FP}$, below which the running is governed by the behavior explained in Sect.~\ref{subsec:nat_log}, rather than by the fixed-point regime.
Linearizing the flow at the NGFP $\lambda_* = 0.1323$, $g_* = 1.0152$, $u_{2*} = 0.0358$, $u_{3*} = -0.4941$ gives the critical exponents $\theta_{1,2} = 2.711 \pm 2.275\,i$ and $\theta_3 = 2.068$, together with an irrelevant direction $\theta_4 = -4.231$. The UV critical surface is therefore three-dimensional, and the corresponding landscape of EFTs is a two-parameter family once one relevant direction is used to set the overall mass unit~\cite{Basile:2021krr}. Because two of the three relevant eigenvalues form a complex-conjugate pair, the associated eigenvectors are complex; we construct a real basis by taking the real eigenvector together with the real and imaginary parts of either eigenvector in the complex-conjugate pair. These three vectors, once normalized, span the tangent space to the critical surface at the fixed point.

We place initial conditions on a sphere of fixed radius $\epsilon=5\cdot10^{-8}$ in tangent space, parametrized by $(\theta,\varphi)$ on a $40\times40$ grid, giving $1600$ starting points at $k_{\rm UV} = 10$ in Planck units. We fix the radius, i.e., the magnitude of the perturbations along the relevant eigendirections, and span the whole space of physically distinct trajectories by sampling the two angles. The radius is kept small so that the starting points lie in the regime where the linearisation defining the critical surface is accurate, and since relevant perturbations grow as $k$ decreases, the trajectories spread out as they are integrated. Each of the $1600$ RG trajectories is integrated until the solver terminates, either because the beta functions develop a singularity or because the system becomes too stiff to continue, and we record the scale at which this happens. We note that the IR-shield mechanism described in Sect.~\ref{subsect:shield} bounds the IR values of the couplings, so that some of them cannot vary across large ranges of values. Different trajectories typically terminate at different scales, but we need to evaluate all of them at a single reference scale to obtain the hypersurface of EFTs from AS at a given sub-Planckian scale. To this end, we select the highest RG scale~$k$, among those at which individual trajectories terminate. For the $R^3$ truncation this gives $\bar{t} = -3.162$, or $\bar{k} \simeq 10^{17}\,\text{GeV}$. Collecting the values $(\lambda, g, u_2, u_3)$ at this scale for all integrated trajectories gives the sampled landscape at $\bar{k} \simeq 10^{17}\,\text{GeV}$.
Since the critical surface is three-dimensional and the coupling space is four-dimensional, the sampled points are bound by a single relation among all couplings, which we now would like to extract. The ideas in the previous subsection and the emerging evidence on the flatness of the AS landscape~\cite{Basile:2021krr,Knorr:2024yiu} motivate us to attempt a fit of the data via linear and quadratic models in $u_2$. Fitting a linear relation
\begin{equation}
  u_3 = \delta - \chi\,\lambda - \rho\, g - \alpha\, u_2
  \label{fitlin}
\end{equation}
by least squares over all sampled trajectories gives
\begin{equation}
u_3 = 0.1195 - 0.2347\,\lambda - 0.5004\,g - 2.3268\,u_2\,,\qquad \text{at }\bar{k} \simeq 10^{17}\,\text{GeV}.
\label{landscape1}
\end{equation}

This relation accounts for $96\%$ of the variance in $u_3$.
To probe the curvature of the surface, we also fit
\begin{equation}
  u_3 = \delta - \chi\,\lambda - \rho\, g - \omega\, u_2 - \alpha\, u_2^2\,,
  \label{fitquad}
\end{equation}
retaining the quadratic term in $u_2$ only, since this is the direction in which a departure from linearity is expected to be largest, and since adding additional quadratic terms would only make the fit better due to a higher number of parameters. The relation between couplings becomes
\begin{equation}
  u_3 = -0.3788 + 0.0656\,\lambda - 0.1656\,g + 2.1660\,u_2- 30.3287\,u_2^{\,2}\,,\qquad\text{at }\bar{k} \simeq 10^{17}\,\text{GeV}\,.
  \label{r3u3quad}
\end{equation}

\begin{figure}[t]
    \centering
    \includegraphics[width=0.496\linewidth]{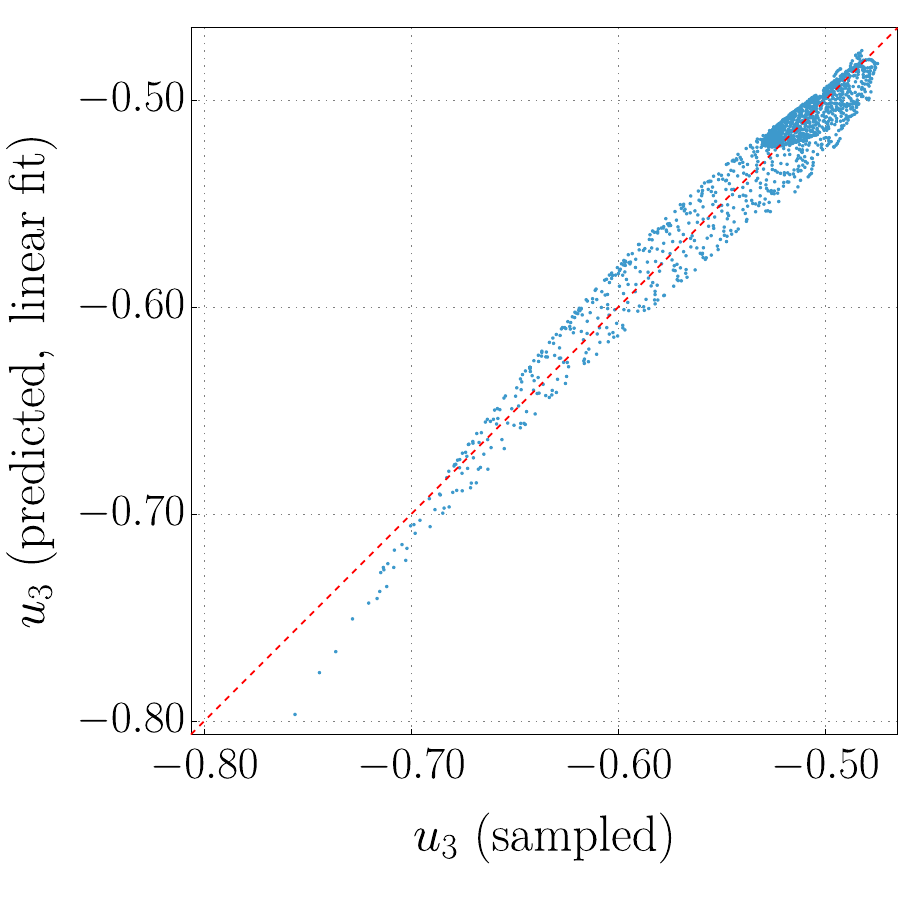}
    \includegraphics[width=0.496\linewidth]{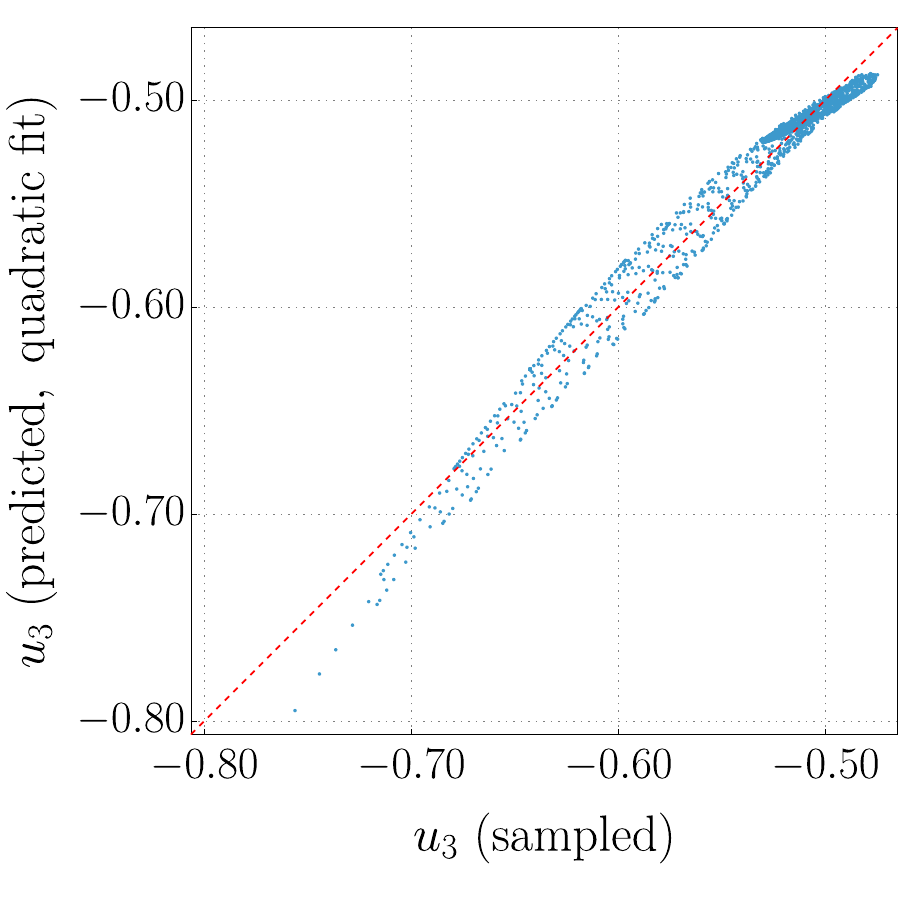}
    \caption{In the truncation including operators up to $R^3$, we show the sampled values of $u_3$ (from the numerical integration with the initial conditions we described) at the common sub-Planckian scale $\bar{k}\simeq10^{17}$\,GeV, plotted against the values predicted by a fit in terms of the remaining couplings. Each point is one of the $1600$ trajectories spanning the UV critical surface and the red dashed line is the $y=x$ line, on which every point would lie if the fitted relation were satisfied exactly. Left panel: linear fit, accounting for $96\%$ of the variance in $u_3$. Right panel: the same data with a quadratic dependence on $u_2$, which reduces the residual variance from $4.1\%$ to $2.7\%$. The residual variance is the fraction of the spread in $u_3$ not reproduced by the fit, and would vanish for a relation satisfied exactly by all trajectories.}
    \label{fits}
\end{figure}

\begin{figure}[t]
    \centering
    \includegraphics[width=0.495\textwidth]{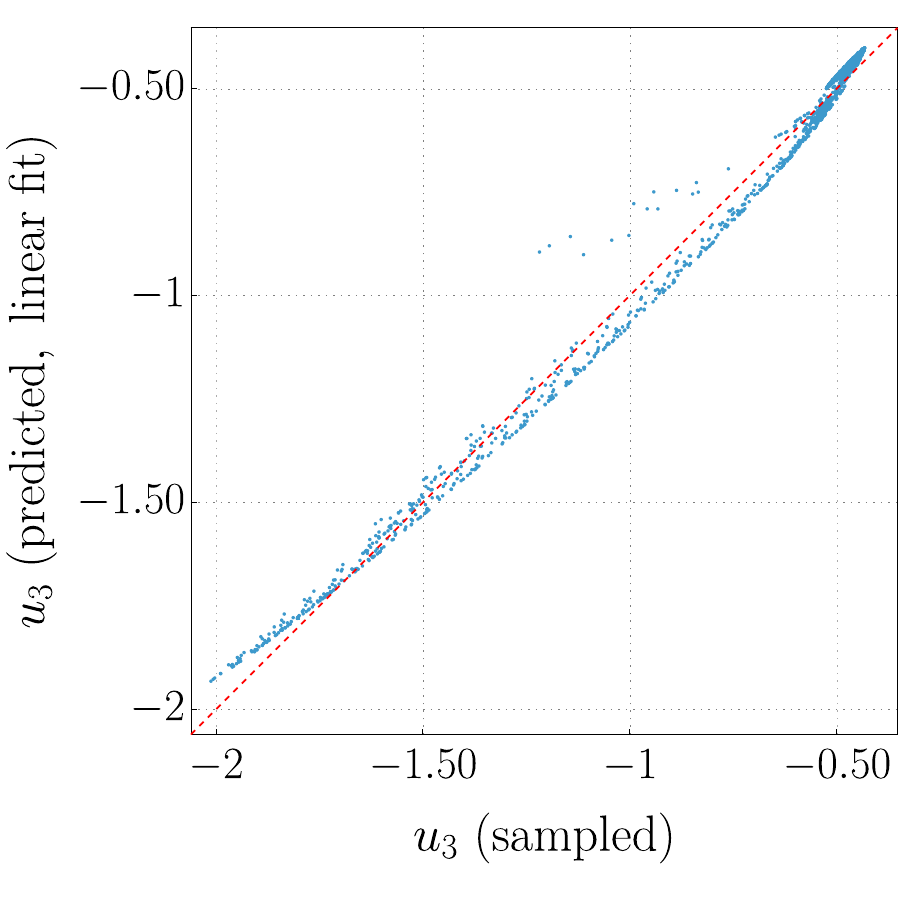}\hfill
  \includegraphics[width=0.495\textwidth]{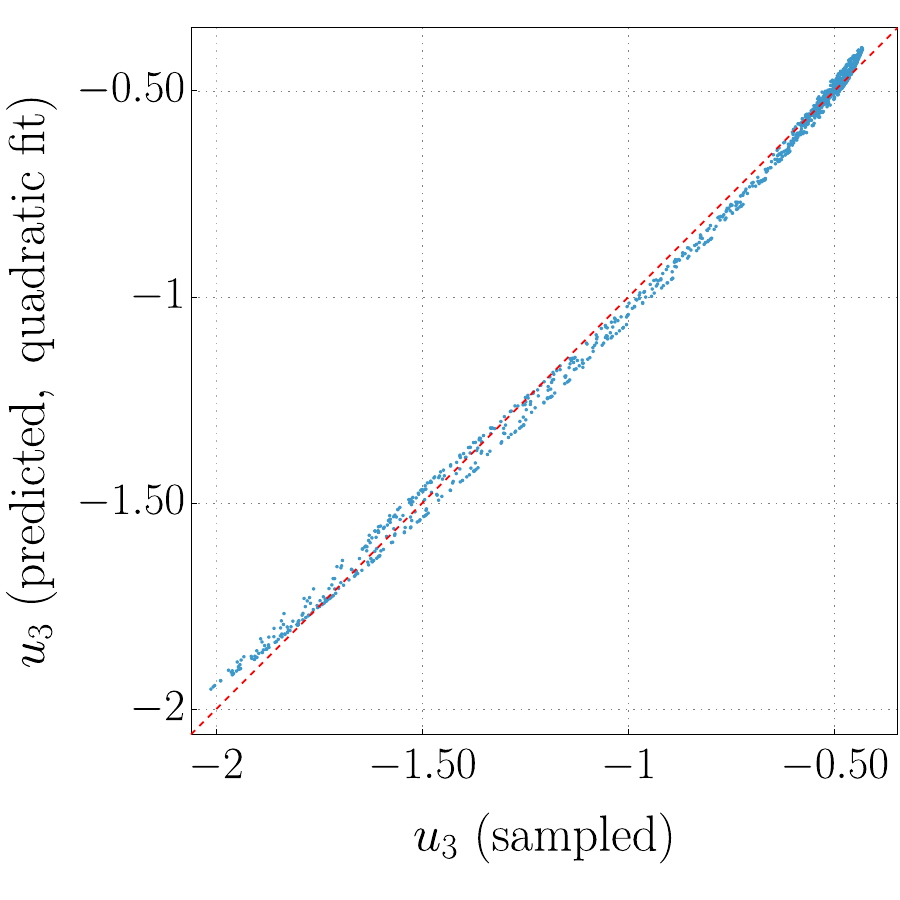}\\[1.5ex]
  \includegraphics[width=0.495\textwidth]{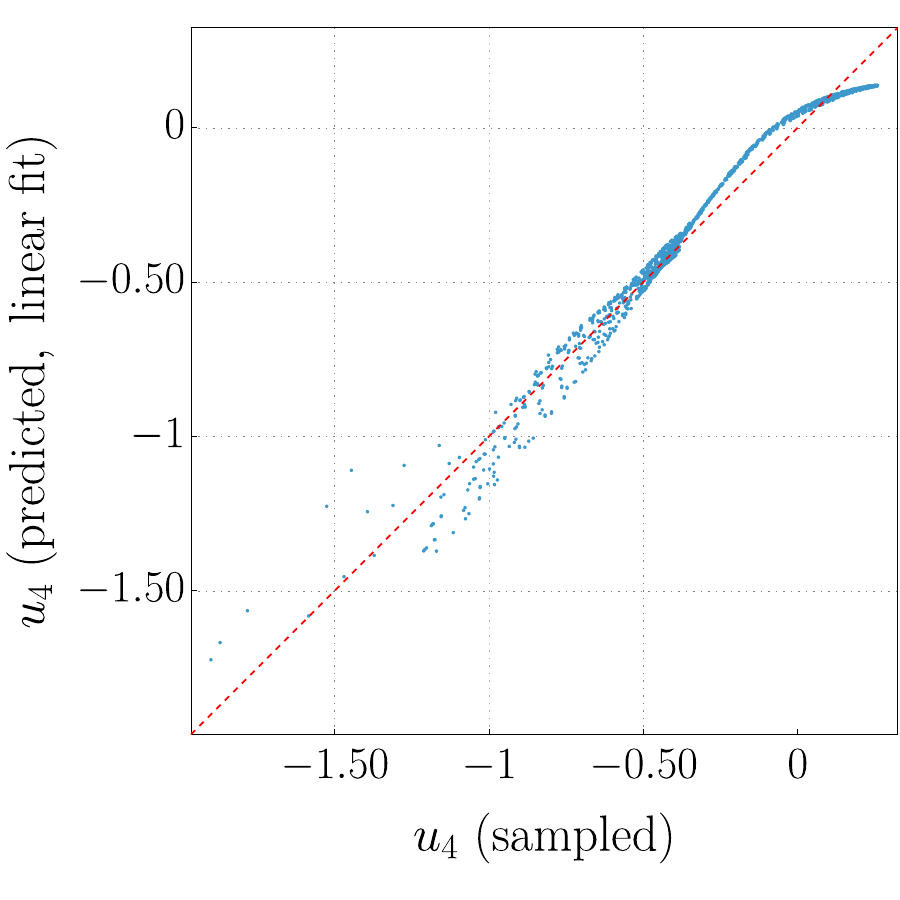}\hfill
  \includegraphics[width=0.495\textwidth]{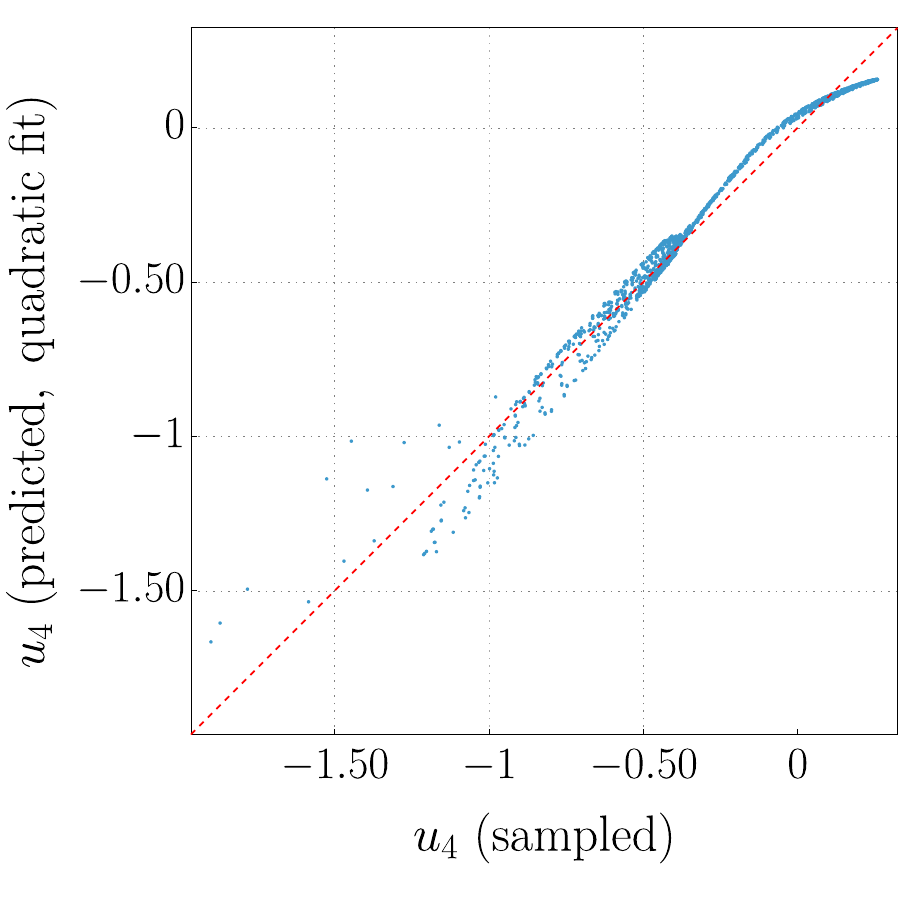}
    \caption{In the truncation including operators up to $R^4$, we show the sampled against predicted values of $u_3$ (top row plots) and $u_4$ (bottom row plots) for the $1600$ asymptotically safe trajectories of the $R^4$ truncation, all evaluated at the common sub-Planckian scale $\bar{t}=-3.447$, or $\bar{k}\simeq7.8\cdot10^{16}\,$GeV. The left column shows the linear fits for $u_3$ and $u_4$. The right column displays the same fits with an additional quadratic term. Each point is the endpoint of a trajectory, and the red dashed line is the $y=x$ line, where each point would be in a perfect fit. The fitted relation accounts for $99.3\%$ (top left panel), $99.7\%$ (top right panel), $97.5\%$ (bottom left panel) and $97.8\%$ (bottom right panel) of the variance of the sampled values.}
    \label{fitsr4}
\end{figure}

The quadratic model reduces the residual variance (meaning the fraction of the spread in $u_3$ not reproduced by the fit, which would vanish for a relation satisfied exactly by all trajectories) from $4.1\%$ to $2.7\%$. However, since adding a parameter always reduces the residual variance, this modest improvement does not justify the extra parameter, and we retain the linear relation, Eq.~\eqref{fitlin}, as the simplest adequate description of the sampled landscape. Both fits are shown in Fig.~\ref{fits}, where the sampled values of $u_3$ are plotted against the values predicted by each model: the points cluster along the $y=x$ line, and the spread is a measure of the residual that the fit does not reproduce. We can read off that the landscape is not perfectly flat, but flatness is still a very good approximation.

The same construction at order $R^4$ still retains a three-dimensional critical surface (the critical exponents being $\theta_{1,2} = 2.864 \pm 2.446\,i$ and $\theta_3 = 1.546$, with two irrelevant directions $\theta_4 = -3.911$ and $\theta_5 = -5.216$), so that there are two irrelevant directions that can be predicted as a function of the three relevant ones. In other words, in this case, the theory space grows to five dimensions and two relations are needed to describe the AS landscape. These are obtained by fitting $u_3$ in terms of $(\lambda,g,u_2)$ and $u_4$ in terms of $(\lambda,g,u_2,u_3)$. All $1600$ trajectories integrate down to sub-Planckian scales, but a larger fraction of them compared to the $R^3$-case terminate before $t=-5$. The common scale thus shifts to $\bar{t} = -3.447$, or $\bar{k}\simeq 7.8\cdot10^{16}\,\text{GeV}$. The sampling procedure is unchanged: we use the same radius $r=5\cdot10^{-8}$ and the same $40\times40$ grid in the tangent space to the critical surface. We also use the same linear (Eq.~\eqref{fitlin}) and quadratic (Eq.~\eqref{fitquad}) fitting models for $u_3$ (now refitted in the enlarged truncation, so that the parameters $(\delta,\chi,\rho,\alpha,\omega)$ take different values from those of the $R^3$ case), and the following linear and quadratic ones for $u_4$:
\begin{align}
    u_4 &= \delta - \chi\lambda - \rho g - \alpha u_2 - \sigma u_3\,,\\
    u_4 &= \delta - \chi\,\lambda - \rho\,g - \alpha\,u_2-\omega\,u_3 - \sigma\,u_3^{\,2}\,.
\end{align}
The two relations between the couplings are, in the linear fit
\begin{align}
    u_3 &= -0.9672 + 1.5317\,\lambda + 0.1613\,g + 5.4284\,u_2\,,
    \label{r4u3linear}\\
    u_4 &= -1.3430 + 1.0300\,\lambda - 0.1702\,g + 13.3035\,u_2- 1.4763\,u_3\,,
    \label{r4u4linear}
\end{align}
whereas in the quadratic fit they read
\begin{align}
    u_3 &= -1.0681 + 1.7923\,\lambda + 0.2354\,g + 7.3678\,u_2-29.6944\,u_2^2\,,
    \label{r4u3quadr}\\
    u_4 &= -1.6936 + 1.3080\,\lambda - 0.1867\,g + 14.2491\,u_2- 2.1442\,u_3 - 0.1535\,u_3^2\,.
    \label{r4u4quadr}
\end{align}
Both fits hold at $\bar{k}\simeq 7.8\cdot10^{16}\,\text{GeV}$ and are shown in Fig.~\ref{fitsr4}. The relation for $u_3$ is well reproduced by a hyperplane, which accounts for $99.3\%$ of the variance of the sampled values, and the quadratic term in $u_2$ improves this to $99.7\%$. The fit for $u_4$ accounts for $97.5\%$. Adding a quadratic term in $u_3$ raises this only to $97.8\%$, and it makes the worst-fitted trajectory even worse, its deviation growing from $0.34$ to $0.44$. As in the cubic case, we therefore use the linear relations in what follows. In absolute terms, the two linear fits are in fact equally good: the largest deviation of any trajectory from the fitted relation is $0.34$ for $u_4$ and $0.32$ for $u_3$. The difference between the two percentages therefore comes from $u_4$ varying less across the sample than $u_3$ does, and not from the relation describing the critical surface less well. Comparing Eq.~\eqref{landscape1} with Eq.~\eqref{r4u3linear}, the coefficients of the $u_3$ relation differ in sign between the two truncations. The individual coefficients are, however, not robust quantities. Repeating the $R^4$ fit at the $R^3$ scale, $\bar{t}=-3.162$, which every $R^4$ trajectory reaches, gives
\begin{equation}
  u_3 = -0.5240 + 0.3438\,\lambda - 0.0941\,g + 2.0930\,u_2 \,,
  \label{r4u3matched}
\end{equation} 
which accounts for $99.8\%$ of the variance and leaves a largest residual of $0.049$, comparable in absolute terms to the $0.041$ of the $R^3$ fit. Between $t=-3.447$ and $t=-3.162$---a change of less than $0.3$ in RG time within the same truncation---the coefficient in front of $u_2$ in the fit changes by a factor of $2.6$ and that in front of $g$ changes sign. The quadratic fits behave the same way, with the coefficient of $u_2^2$ taking the values $-29.69$ and $-17.19$ at the two scales. Overall, our computations seem to indicate that at any fixed Planckian or sub-Planckian scale $\bar{k}$, the relations among the couplings are approximately linear, with the specific coefficients depending on the scale $\bar{k}$ at which the intermediate set of EFTs from AS is computed. The relations we derived are the output of the first step of our patching strategy; we will use them as an input in Sect.~\ref{subsec:nat_full} to further evolve the couplings according to their IR running, from $\bar{k}$ to $k_\text{infl}$. This will allow us to get a final assessment of the naturalness of Starobinsky inflation. Before doing so, in Sect.~\ref{subsec:nat_canonical} and Sect.~\ref{subsec:nat_log}, we discuss two aspects of the UV-to-IR running that can influence the naturalness question: the interplay between geometry of the landscape and canonical scaling, and the potential impact of logarithmic corrections.

\subsection{Interlude: interplay between geometry of the AS landscape, naturalness, and canonical scaling \label{subsec:nat_canonical}}

The analysis of Sect.~\ref{FP&geom} shows that if the landscape is flat, the higher-curvature couplings are strongly suppressed at inflationary scales, and in particular their values lie many orders of magnitude below the observational bounds~\eqref{eq:unn-val}. In this section we ask what is the mechanism responsible for this suppression, thus providing a conceptual understanding of why the geometry of the AS landscape matters.

A natural candidate is canonical scaling, which we have indeed used in Sect.~\ref{FP&geom} to map the couplings at the fixed-point and inflationary scales. In what follows, we show explicitly how this mechanism works. At the transition scale $k_{\rm FP}\simeq M_{\rm Pl}$  the flow leaves the fixed-point regime, so that between a sub-Planckian scale $\bar{k}\lesssim k_{\rm FP}$ and the inflationary scale $k_{\rm infl}$, the couplings are small and their running is expected to be controlled by the canonical dimensions of the operators,
\begin{equation}
  u_i(k) \simeq u_i(\bar{k}) \left(\frac{k}{\bar{k}}\right)^{2i-2}\,, \qquad i=2,3,4,\dots
  \label{eq:canonical_scaling}
\end{equation}
Operators with $i\geq 3$ are irrelevant, so the corresponding dimensionless couplings are driven to zero as $k$ decreases, and the more so the higher the order of the operator. For $u_3$ and $u_4$ this gives
\begin{equation}
  u_3(k_{\rm infl}) = u_3(\bar{k})\,x^{-4}\,, \qquad
  u_4(k_{\rm infl}) = u_4(\bar{k})\,x^{-6}\,, \qquad
  x \equiv \frac{\bar{k}}{k_{\rm infl}} \sim 10^{6}\,.
  \label{eq:canonical_suppression}
\end{equation}
Even values of order one at $\bar{k}$ are thus reduced by several orders of magnitude, with a factor $10^{-24}$ for $u_3$ and $10^{-36}$ for $u_4$. The orders of magnitude found in Sect.~\ref{FP&geom} using a fixed-point expansion of all running couplings are compatible with this picture, which suggests that the suppression originates in the large hierarchy between $\bar{k}$ and $k_{\rm infl}$, combined with the irrelevance of the higher-order operators with respect to the AS fixed point. The relevant couplings $\lambda$, $g$ and $u_2$ are not suppressed in this way. The mechanism requires, however, that $u_3(\bar{k})$ and $u_4(\bar{k})$ are themselves $\sim\mathcal{O}(1)$, that is, that they are related to the relevant couplings with $\mathcal{O}(1)$ coefficients. This is where the geometry of the UV-critical surface enters: If $u_3(\bar{k})$ is linearly related to the relevant couplings, Eq.~\eqref{eq:canonical_suppression} yields the required suppression. If instead the relation were quadratic, i.e., if $u_3 \sim u_2^2$ along the trajectories, then $u_3(\bar{k}) \sim u_2^2(\bar{k}) \sim u_2^2(k_{\rm infl})\,x^{4}$, and the factor $x^{-4}$ cancels exactly the suppression factor in Eq.~\eqref{eq:canonical_suppression}. This yields $u_3(k_{\rm infl}) = \mathcal{O}(1)$, in tension with the bound~\eqref{eq:unn-val}. This line of arguments explains why and how the interplay between geometry of the AS landscape and canonical scaling is essential to determine the naturalness of Starobinsky inflation. 

This explanation rests, however, on the assumption that the running between $\bar{k}$ and $k_{\rm infl}$ is purely canonical. As discussed in Sect.~\ref{subsec:IR_features}, this is not the case already for the linearized flow around the GFP, in which mixings between different scalings can occur. While we will see that this does not impact the arguments in this section, another ingredient in principle could: the logarithmic running of non-marginal couplings. We discuss this in the next subsection.

\subsection{Interlude: impact of logarithmic corrections on the naturalness question} \label{subsec:nat_log}

Canonical scaling accounts only for the classical mass dimensions of the couplings. The actual RG flow also receives quantum contributions, and in particular the fact that all couplings can enter each beta function, generating a logarithmic running also for non-marginal couplings~\cite{Knorr:2024yiu,DelPorro:2025fiu}. This comes on top of the non-trivial mixings between different canonical scalings discussed in Sect.~\ref{subsec:IR_features}. With the goal of understanding which ingredients can favor or disfavor the naturalness of Starobinsky inflation, in this section we investigate  the impact of logarithmic corrections to the higher-curvature couplings on the canonical suppression of couplings detailed in the previous section.

To assess their potential impact, we introduce a phenomenological parametrization of logarithmic corrections to the running of the higher-curvature couplings $u_i$, with generic coefficients $b_i$,
\begin{equation}
  \begin{aligned}
    \lambda(\bar{k}) &= \lambda(k_{\rm infl})\,x^{-2}\,, \qquad g(\bar{k}) = g(k_{\rm infl})\,x^{2}\,,\\
    u_i(\bar{k}) &= \big[\,u_i(k_{\rm infl}) + b_i \log x\,\big]\,x^{2i-2}\,, \qquad i=2,3,4\,,
  \end{aligned}
  \label{eq:generic_log}
\end{equation}
where $x=\bar{k}/k_{\rm infl}$. For $b_i=0$, these expressions describe the canonical scalings of all couplings. The values of the coefficients $b_i$ must ultimately be determined from the RG flow. We first consider the cubic truncation.  If, as a simple example, we use the linear relation~\eqref{eq:tangent_cubic_b} and once again the observational values~\eqref{eq:obs-values} for the relevant dimensionless couplings, we find
\begin{equation}
  u_3(k_{\rm infl}) \simeq -1.9\times 10^{-13} + 1.6\times 10^{-11}\, b_2 - 13.8\, b_3\,.
  \label{eq:u3_generic_b}
\end{equation}
The first term corresponds to the canonical result. The contribution
proportional to $b_2$ is suppressed by $x^{-2}$ and is therefore
strongly suppressed for the scale hierarchy considered here. By
contrast, the contribution proportional to $b_3$ is not accompanied
by an inverse power of $x$: it is enhanced by the factor
$\log x\simeq13.8$.

Consequently, even a logarithmic correction that is modest at the
level of the RG flow can dominate over the canonically suppressed
contribution to $u_3(k_{\rm infl})$. In the absence of cancellations with the other terms, values of $|b_3|$ of order $10^{-5}$ or larger would make $|u_3(k_{\rm infl})|$ exceed the observational bound $|u_3|\lesssim10^{-4}$. This illustrates how logarithmic corrections could potentially spoil the hierarchy required for naturalness.

We now extend this analysis to the quartic truncation and we obtain
\begin{equation}
  \begin{aligned}
    u_3(k_{\rm infl}) &\simeq
-1.88\times10^{-13}
+1.56\times10^{-11}b_2
-13.8b_3\,,\\
u_4(k_{\rm infl}) &\simeq
-2.10\times10^{-25}
+1.74\times10^{-23}b_2
-13.8b_4\,.
  \end{aligned}
\end{equation}
As in the cubic truncation, the contributions proportional to $b_2$ are strongly suppressed by the scale hierarchy, whereas those proportional to $b_3$ and $b_4$ are not. Thus, logarithmic corrections to the running of either higher-curvature coupling can potentially overcome the canonical suppression.

The coefficients $b_i$, however, are not arbitrary parameters of the physical RG flow: their values and the actual impact of the logarithmic terms must be determined from the RG equations. We address this in the following section by replacing the parametrization above with the actual IR running derived in Sect.~\ref{subsec:IR_beyond}.

\subsection{All together: matching with the IR running and naturalness}
\label{subsec:nat_full}

In Sect.~\ref{subsec:landscape} we performed a full-fledged integration of the RG flow, from the UV fixed point down to the sub-Planckian scale $\bar{k}$. At this scale, we discovered the irrelevant couplings to be linearly related to the relevant ones: the intermediate set of EFTs is an approximately flat hypersurface in theory space, described by Eq.~\eqref{landscape1} for a cubic truncation of the flow, and by Eqs.~\eqref{r4u3linear} and~\eqref{r4u4linear} for a quartic truncation. Based on the toy models in Sect.~\ref{FP&geom} and on the arguments in Sect.~\ref{subsec:nat_canonical}, the flatness at sub-Planckian scales, when matched with the canonical scaling regime, can contribute to making Starobinsky inflation natural. This, however, depends on the specific scaling of couplings about the GFP, and also on how logarithmic corrections enter (cf.~\ref{subsec:nat_log}). In this section we combine the results of the first part of the patching strategy (cf. Sect.~\ref{subsec:landscape}) with the full IR scaling derived in Sect.~\ref{subsec:IR_beyond} and the observational values~\eqref{eq:obs-values} for the relevant dimensionless couplings to get a final result on the naturalness of Starobinsky inflation.

The resulting matching relations depend on the behavior of the dimensionful cosmological constant in the IR. In the following, we focus on the separatrix, with vanishing dimensionful cosmological constant, which is the case relevant for our setup, as discussed in Sect.~\ref{subsec:IR_beyond}.

In the cubic truncation, using the linear fit of Eq.~\eqref{landscape1} for the landscape at $k=\bar{k}$ and Eqs.~\eqref{eq:runninglgCC0}--\eqref{eq:runningu3GFP} for the full-fledged IR running, we obtain
\begin{equation}
  u_3(k_{\rm infl}) \simeq 5.5\times10^{-13}\,.
\end{equation}
This value is of the same order as the canonical result of Sect.~\ref{subsec:nat_log} and many orders of magnitude below the bound $|u_3|\lesssim10^{-4}$: since $u_3$ does not acquire a logarithm in the IR solution, its canonical suppression is not affected.

In the quartic truncation we use for the landscape at $k=\bar{k}$ the two linear relations of Sect.~\ref{subsec:landscape}, Eqs.~\eqref{r4u3linear} and~\eqref{r4u4linear}, which express $u_3$ in terms of $(\lambda,g,u_2)$ and $u_4$ in terms of $(\lambda,g,u_2,u_3)$. Matching them with the full-fledged IR running~\eqref{eq:runninglgCC0}--\eqref{eq:runningU4CC0}, we obtain 
\begin{equation}
  u_3(k_{\rm infl}) \simeq -7.5\times10^{-13}\,, \qquad
  u_4(k_{\rm infl}) \simeq -6.1\times10^{-14}\,.
\end{equation}
Both couplings are strongly suppressed, and lie many orders of magnitude below the observational bounds~\eqref{eq:bounds}, so that the naturalness of Starobinsky inflation is preserved also in this case.

Overall, for the case $\Lambda_0=0$ and within the linear description of the landscape, the full IR running does not spoil the hierarchy required by the Starobinsky regime. Logarithmic contributions to the RG running could in principle affect the suppression of higher-curvature couplings, but their physical impact depends on their origin: logarithms associated with marginal couplings differ from those generated by the mixing between couplings. As discussed in~\cite{Knorr:2026vax}, once matter fluctuations are taken into account, logarithmic contributions arising from coupling mixing may be suppressed by factors involving the ratio $M/M_{\rm Pl}$, where $M$ denotes the mass of a matter field. This implies that even in the case higher-derivative contributions beyond those we have studied would introduce logarithmic running dominating over the canonical-scaling suppression, the inclusion of matter fluctuations would restore the naturalness again.

We finally translate our results to Lorentzian settings by Wick rotating as explained in Sect.~\ref{sect:setup}, $c_i = -u_i(k_{\rm infl})$. In this case, our linear-fit results with $\Lambda_0=0$ give $c_3 \simeq -5.5\times10^{-13}$ in the cubic truncation and $c_3 \simeq +7.5\times10^{-13}$, $c_4 \simeq +6.1\times10^{-14}$ in the quartic one. The sign change of $c_3$ between our cubic and quartic truncations mirrors the one found in the literature on observational constraints, which in general does not agree on the sign. Specifically, some studies of truncations up to cubic order constrained with Planck data assume a positive $c_3$ from the start, for stability reasons~\cite{Ivanov:2021chn}, while the inclusion of ACT data seems to favor a small negative value~\cite{Bezerra-Sobrinho:2025gfg}. An extension up to $R^5$ instead yields $c_3>0$ and $c_4<0$~\cite{Ketov:2025cqg}. While our finding of different coefficient signs in different truncations seems in qualitative agreement with these studies, it must be noted that our coefficients lie far below current data sensitivity. Nevertheless, the fact that the sign is not firmly established by observations motivates the focus of the present work on the magnitude of the coefficients only, which is well within the bounds~\eqref{eq:cheong-rel} and~\eqref{eq:unn-val} and hence compatible with a natural regime of Starobinsky inflation.

\section{Conclusions}
\label{sect:conclusions}

In this work, we have revisited the naturalness of Starobinsky inflation from the perspective of effective field theory (EFT) and renormalization-group (RG) dynamics. When embedded in a general gravitational EFT, realizing Starobinsky inflation without spoiling the data requires a specific hierarchy among the coefficients of the higher-curvature operators: while the coupling multiplying the $R^2$ term must be sufficiently large to reproduce the observed amplitude of primordial scalar fluctuations, the couplings associated with higher-order curvature invariants must remain sufficiently suppressed. This hierarchy is not generic from the viewpoint of EFT power counting, and hence appears to be unnatural from an RG perspective. Our aim has been to make the question of naturalness concrete and quantitative by investigating whether and how such a hierarchy can(not) emerge dynamically from a consistent RG flow.

To this end, we have used asymptotically safe (AS) gravity as a framework in which RG trajectories are well-defined across all scales, and in which most of the higher-derivative couplings (the irrelevant ones) can be predicted as a function of the free parameters of the theory: the cosmological constant, Newton's and $R^2$ couplings. We have constructed the full RG evolution of the gravitational couplings up to and including operators of order $R^4$, considering a sample of trajectories departing from the ultraviolet (UV) fixed point and homogeneously spanning the UV critical surface. In this context, it is the requirement of UV completeness that induces specific relations among the Wilson coefficients of the theory, and that allows one to predict the irrelevant ones in terms of the relevant couplings. A full numerical integration of the RG flow, from the UV down to inflationary scales, proved challenging.  Hence, we have adopted a two-step strategy, first integrating the flow from the UV down to sub-Planckian scales, and then using the infrared (IR), canonical scaling of couplings, to obtain predictions at inflationary scales. A central finding of the first step is that the landscape of EFTs stemming from AS has an approximately flat geometry: the relations among the couplings are approximately linear, and this turns out to be a crucial ingredient in the naturalness question. Moreover, the approximate flatness of the landscape is a recurrent finding in the recent literature of AS~\cite{Basile:2021krr,Knorr:2024yiu}, making it an emerging feature of AS. 

The linear relations found from the numerical integration provide the starting point for the subsequent RG evolution towards inflationary scales, where the hierarchy required by cosmological observations is to be assessed. Two ingredients turned out to be essential to characterize this part of the RG evolution and the naturalness question. The first is the canonical scaling of the couplings in the vicinity of the GFP. In particular, canonical scaling relates the values of the couplings at inflationary scales, $k_{\rm infl}$, to those at sub-Planckian scales, $\bar{k}$, through powers of the dimensionless ratio $x=\bar{k}/k_{\rm infl}$, with the appropriate exponent determined by the canonical mass dimension of each coupling. Consequently, even relatively mild differences in the scaling dimensions of different operators can translate into substantial hierarchies at the much lower inflationary scale. The second ingredient is the logarithmic running of the couplings, which provides corrections to this canonical scaling and can potentially modify the resulting hierarchy.

The interplay between these effects, together with the approximately flat structure of the sub-Planckian AS landscape and the fact that the UV critical surface has three relevant directions, leads to a remarkable result. The higher-curvature couplings obtained at inflationary scales naturally satisfy the hierarchy required by cosmological observations. In other words, the suppression of higher-order operators relative to the quadratic-curvature contribution does not need to be imposed as an independent assumption: it can emerge from the RG evolution of the theory. Within this framework, an approximate regime of Starobinsky inflation becomes natural from the RG perspective, despite appearing highly tuned when considered solely as a particular truncation of the gravitational EFT. 

There are, however, several subtleties in the RG evolution that deserve particular attention. First, the scaling around the GFP is not purely canonical: the stability matrix contains off-diagonal contributions, which mix the couplings, and a non-trivial Jordan block, which produces a logarithmic correction to the running of $u_2$. We have explicitly accounted for these effects. The mixing contributions do not spoil the naturalness mechanism identified above, and the resulting hierarchy remains robust against their inclusion. The logarithm in $u_2$ affects only the coefficient of the $R^2$ term itself, and not the suppression of higher-order operators. Logarithmic corrections to the higher-order couplings would instead be more delicate: they can compete with the canonical scaling and undermine the suppression of the corresponding operators. Such a correction does arise for $u_4$, but only in the deep IR, which lies far below the scales relevant for inflation for a sufficiently small cosmological constant. Since a small cosmological constant is itself a phenomenologically relevant requirement, this observation establishes a non-trivial connection between the smallness of $\Lambda$ and the naturalness of Starobinsky inflation. Thus, the same condition that selects phenomenologically interesting trajectories also ensures that the logarithmic corrections do not invalidate the hierarchy required by inflationary observations. This points towards a deeper interplay between the cosmological constant problem and the naturalness of inflationary dynamics.

All in all, our conclusions on the naturalness of Starobinsky inflation rely on several interconnected ingredients: the flatness of the AS landscape that makes dimensionless couplings of the same order at sub-Planckian scales, the existence of three relevant directions coinciding with the three lowest-order couplings, the smallness of the cosmological constant, and the canonical scaling of couplings, which provides suppression factors for all irrelevant couplings, making them highly suppressed at inflationary scales with respect to the $R^2$-coupling. The absence of any of these ingredients would make Starobinsky inflation unnatural.

Beyond its implications for Starobinsky inflation, our analysis has also revealed several features of the RG flow of $f(R)$ theories that may be relevant for AS in general: 
\begin{itemize}
    \item \textit{Flatness of the AS landscape}: our work provides further evidence that the hypersurface of EFTs stemming from an AS fixed point is approximately flat, implying that the irrelevant couplings in the IR are related to relevant ones via approximately linear relations. 
    \item \textit{IR shields}: higher-derivative truncations can exhibit multiple IR fixed points, which act as effective IR shields for the RG trajectories. Such fixed points constrain the evolution of the relevant couplings and can restrict the range of values that these couplings attain in the IR. In this sense, IR fixed points can enhance the predictive power of the system by limiting the region of the effective theory space accessible to RG trajectories. This observation suggests that the structure of the IR flow deserves as much attention as the UV fixed-point properties when assessing the predictive content of asymptotically safe models.
    \item \textit{(Non-)canonical scalings}: While Gaussian fixed points are often associated with straightforward canonical scaling, their stability matrices can exhibit non-trivial features, in particular the corresponding eigenvectors may not always be aligned with the axes defined by the individual couplings, or the eigenvectors may not form a complete basis. In such cases, the canonical scaling of the eigenperturbations translates into a non-trivial mixing of the scaling behavior of individual couplings. This structure can be encoded in Jordan blocks of the stability matrix and gives rise to logarithmic contributions in the RG evolution, including for non-marginal couplings. Consequently, the canonical dimensions of individual operators do not necessarily provide a complete description of their scaling behavior, even in the vicinity of a Gaussian fixed point.
    \item \textit{Cosmological constant and logarithmic running}: Depending on whether $\Lambda$ vanishes exactly or is non-zero, however small, the existence and structure of logarithmic form factors in certain couplings can change. In other words, the limit of vanishing cosmological constant need not be equivalent to considering a theory with an arbitrarily small but finite cosmological constant.
\end{itemize}

To summarise, our findings provide a concrete RG-based mechanism through which the hierarchy required by Starobinsky inflation can emerge naturally. This comes from a combination of non-trivial ingredients, from the approximately flat structure of the AS landscape at sub-Planckian scales, to the canonical scaling towards the inflationary regime, and the absence of dangerous logarithmic corrections along phenomenologically relevant trajectories. At the same time, the analysis exposes a number of potentially important features of higher-derivative RG flows, including the role of IR fixed points, non-trivial scalings around Gaussian fixed points, and the sensitivity of logarithmic form factors to the cosmological constant. Taken together, these results suggest that the interplay between UV predictivity and IR dynamics may provide a general mechanism for understanding apparently unnatural hierarchies in gravitational EFTs. 

\acknowledgments

The authors thank F. Del Porro for discussions and M. Pauly for early contributions to this work.  The research of A.P. and F.F. is supported by a research grant (VIL60819) from VILLUM FONDEN. The Center of Gravity is a Center of Excellence funded by the Danish National Research Foundation under grant No. 184. This project has received funding from the European Union's Horizon 2020 research and innovation programme under the Marie Sklodowska-Curie grant agreement No 101299389 and No 101131233. This work is supported by Simons Foundation International \cite{sfi} and the Simons Foundation \cite{sf} through Simons Foundation grant SFI-MPS-BH-00012593-11. S.G. acknowledges support from the Simons Investigator award SFI-MPS-SIP-00017355, thanks NBIA for hospitality, and acknowledges funding from the Rosenfeld Visiting Grant and the Center of Gravity for two research visits where part of this work was completed. F.F. is grateful to the Max Planck Institute for Physics in Garching, where part of the project was developed, and to Perimeter Institute for Theoretical Physics and Harvard University for their hospitality during the completion of this work. Research at Perimeter Institute is supported in part by the Government of Canada through the Department of Innovation, Science and Economic Development Canada and by the Province of Ontario through the Ministry of Colleges and Universities.

\printbibliography{}

\end{document}